\documentclass[12pt]{article}
\usepackage{array,amsmath,amssymb,epsfig,verbatim}
\begin{document}


\def\slash#1{\ooalign{$\hfil/\hfil$\crcr$#1$}}

\newcommand{\kIR}{K_{\Lambda\infty}}
\newcommand{\kUV}{K_{0\Lambda}}
\newcommand{\Str}{\mbox{STr}}
\newcommand{\Res}{\mbox{Res}}
\newcommand{\LdL}{\Lambda\partial_\Lambda}

\newcommand{\equal}[1]
{
\buildrel{#1}\over=
}

\def\ap#1#2#3{Ann.\ Phys.\ (NY) #1 (19#3) #2}
\def\cmp#1#2#3{Commun.\ Math.\ Phys.\ #1 (19#3) #2}
\def\ib#1#2#3{ibid.\ #1 (19#3) #2}
\def\zp#1#2#3{Z.\ Phys.\ #1 (19#3) #2}
\def\np#1#2#3{Nucl.\ Phys.\ B#1 (19#3) #2}
\def\pl#1#2#3{Phys.\ Lett.\ #1B (19#3) #2}
\def\pr#1#2#3{Phys.\ Rev.\ D #1 (19#3) #2}
\def\prb#1#2#3{Phys.\ Rev.\ B #1 (19#3) #2}
\def\prep#1#2#3{Phys.\ Rep.\ #1 (19#3) #2}
\def\prl#1#2#3{Phys.\ Rev.\ Lett.\ #1 (19#3) #2}
\def\rmp#1#2#3{Rev.\ Mod.\ Phys.\ #1 (19#3) #2}

\newcommand{\figura}[3]
{
\begin{figure}[t]
  \begin{center}
  \mbox{\epsfig{file=#1,height=#2}}
  \end{center}
  \caption{{{\small #3}}}
\end{figure}
}
\newcommand{\figuraX}[2]
{
\begin{figure}[h]
  \begin{center}
  \mbox{\epsfig{file=#1,height=#2}}
  \end{center}
\end{figure}
}
\renewcommand{\div}
{
\mbox{div}
}
\newcommand{\binomio}[2]
{
\left(\begin{array}{ccc}#1\cr #2\end{array}\right)
}
\renewcommand{\pmatrix}[1]
{
\left(\begin{array}{ccc}#1\end{array}\right)
}
\newcommand{\disegno}[2]
{
\begin{center}                             
  \begin{picture}(#1) #2
  \end{picture}
\end{center}
}
\newcommand{\des}[1]
{
\overleftarrow{#1}
}
\newcommand{\ddes}[2]
{
{\overleftarrow\delta #1\over\delta #2}
}
\newcommand{\dsin}[2]
{
{\overrightarrow\delta #1\over\delta #2}
}
\newcommand{\iacc}{\`\i{\ }}
\renewcommand{\a}{{\tilde a}} \renewcommand{\b}{{\tilde b}}
\renewcommand{\c}{{\tilde c}} \renewcommand{\d}{{\tilde d}}
\newcommand{\e}{{\tilde e}}\newcommand{\f}{{\tilde f}}
\newcommand{\g}{{\tilde g}}\newcommand{\h}{{\tilde h}}
\newcommand{\ii}{{\tilde \imath}} \newcommand{\jj}{{\tilde \jmath}}
\renewcommand{\k}{{\tilde k}} \renewcommand{\l}{{\tilde l}}
\newcommand{\m}{{\tilde m}} \newcommand{\n}{{\tilde n}}
\renewcommand{\o}{{\tilde o}} \newcommand{\p}{{\tilde p}}
\newcommand{\q}{{\tilde q}} \renewcommand{\r}{{\tilde r}}
\newcommand{\s}{{\tilde s}} \renewcommand{\t}{{\tilde t}}
\renewcommand{\u}{{\tilde u}} \renewcommand{\v}{{\tilde v}}
\newcommand{\w}{{\tilde w}} \newcommand{\x}{{\tilde x}}
\newcommand{\y}{{\tilde y}} \newcommand{\z}{{\tilde z}}
\newcommand{\ux}{{\underline x}} \newcommand{\uy}{{\underline y}}
\newcommand{\uv}{{\underline v}} \newcommand{\uw}{{\underline w}}
\newcommand{\uz}{{\underline z}}
\newcommand{\bi}{\bar\imath} \newcommand{\bj}{\bar\jmath}


\newcommand{\Tr}
  {\mbox{Tr}}
\newcommand{\STr}
  {\mbox{STr}}
\newcommand{\A}
  {{\cal A}}
\newcommand{\B}
  {{\cal B}}
\newcommand{\C}
  {{\cal C}}                                  
\newcommand{\D}
  {{\cal D}}
\newcommand{\De}
  {{\cal D}}
\newcommand{\E}
  {{\cal E}}
\newcommand{\F}
  {{\cal F}}
\newcommand{\G}
  {{\cal G}}
\renewcommand{\H}
  {{\cal H}}
\newcommand{\I}
  {{\cal I}}
\newcommand{\J}
  {{\cal J}}
\newcommand{\K}
  {{\cal K}}
\renewcommand{\L}
  {{\cal L}}
\newcommand{\M}
  {{\cal M}}
\newcommand{\N}
  {{\cal N}}
\renewcommand{\O}
  {{\cal O}}
\renewcommand{\P}
  {{\cal P}}
\newcommand{\Q}
  {{\cal Q}}
\newcommand{\R}
  {{\cal R}}
\renewcommand{\S}
  {{\cal S}}
\newcommand{\T}
  {{\cal T}}
\newcommand{\U}
  {{\cal U}}
\newcommand{\V}
  {{\cal V}}
\newcommand{\W}
  {{\cal W}}
\newcommand{\X}
  {{\cal X}}
\newcommand{\Y}
  {{\cal Y}}
\newcommand{\Z}
  {{\cal Z}}
\newcommand{\uno}
  {\unity}
\newcommand{\ep}
  {\varepsilon}
\newcommand{\lie}
  {\ell_\varepsilon}
\newcommand{\lt}[1]
  { \ell_{t_{#1}} }
\newcommand{\leps}[1]
  { \ell_{\varepsilon_{#1}} }
\newcommand{\KG}
{\mbox{ \tiny \begin{tabular}{|c|} \hline \\ \hline \end{tabular} } }
\newcommand{\sedici}
{\overline{16}}

\newcommand{\ansatz}{{\it Ans\"atze }}


\newcommand{\PropI}
{
\Delta^{-1}_{\Lambda\Lambda_0}
}
\newcommand{\Prop}
{
\Delta_{\Lambda\Lambda_0}
}
\newcommand{\rif}[1]
  {(\ref{#1})}

\newcommand{\formula}[2]
  { \begin{equation} \label{#1} #2 \end{equation} }
\newcommand{\formule}[2]
  { \begin{align} \label{#1} #2 \end{align} }
\newcommand{\formulona}[2]
{
  \begin{equation} \label{#1}
  \begin{split}
    #2 
  \end{split}
  \end{equation}
}
\newcommand{\formulonaX}[1]
{
  \begin{equation} 
    \begin{split}
      #1
    \end{split}
  \end{equation}
}
\newcommand{\graffa}[2]
{
  \begin{equation} \label{#1}
    \left\{ \begin{array}{lll} #2 \end{array} \right.
  \end{equation}
}
\newcommand{\graffaetichettata}[3]
{
  \begin{equation} \label{#2}
    #1\ \left\{ \begin{array}{lll} #3 \end{array} \right.
  \end{equation}
}
\newcommand{\formulaX}[1]
{  \begin{equation} #1 \end{equation} 
}
\newcommand{\formuleX}[1]
{ 
  \begin{align} #1 \end{align} 
}
\newcommand{\graff}[1]
{
  $$  \left\{ \begin{array}{lll} #1 \end{array} \right. $$
}
\newcommand{\graffet}[2]
{
 $$  #1\ \left\{ \begin{array}{lll} #2 \end{array} \right. $$
}
\newcommand{\graffaX}[1]
{
  \begin{equation} 
    \left\{ \begin{array}{lll} #1 \end{array} \right.
  \end{equation}
}
\newcommand{\graffaetichettataX}[2]
{
  \begin{equation}
    #1\ \left\{ \begin{array}{lll} #2 \end{array} \right.
  \end{equation}
}
\newcommand{\tabella}[4]
{
    \label{#1}
    \begin{center}
      #2\\ $\ $ \\
      \begin{tabular}{#3}
        #4  
      \end{tabular}
    \end{center}
}
\newcommand{\tabarray}[4]
{
    \label{#1}
    \begin{center}
      #2 $$ \begin{array}{#3} #4 \end{array} $$
    \end{center}
}
\newcommand{\tabellaX}[3]
{
  \begin{table}
   \caption{#1}
    \begin{center}
      \begin{tabular}{#2}#3
      \end{tabular}
    \end{center}
  \end{table}
}
\newcommand{\tabarrayX}[4]
{
  \begin{table}
    \caption{#2}
    \label{#1}
    \begin{center}
       $$ \begin{array}{#3} #4 \end{array} $$
    \end{center}
  \end{table}
}
\newcommand{\detizero}[1]
{
\left.{d\over dt}#1 \right|_{t=0}
}
\newcommand{\prendi}[2]
{
\left. #1 \right|_{#2}
}
\newcommand{\dezero}[2]
{
\left.{\mbox{d} #1\over \mbox{d} #2} \right|_{#2=0}
}
\newcommand{\deltazer}[2]
{
\left.{\delta#1\over\delta#2}\right|_{\breve \mu=0}
}
\newcommand{\deltazero}[2]
{
\left.{\delta #1\over \delta #2} \right|_{#2=0}
}
\newcommand{\desude}[1]
{
{\partial \over \partial #1}
}
\newcommand{\dxx}[2]
{
{ \partial x^{#1}\over\partial x^{#2} }
}
\newcommand{\dd}[2]
{
\frac{\delta{#1}}{\delta{#2}}
}
\newcommand{\dede}[2]
{
{\partial{#1}\over\partial{#2}}
}
\newcommand{\dedi}[2]
{
{\mbox{d}{#1}\over \mbox{d}{#2}}
}

	
\begin{titlepage}
\renewcommand{\thefootnote}{\fnsymbol{footnote}}
\begin{flushright}
     LPTHE 00-18\\
     May 2000 \\
\end{flushright}
\par \vskip 10mm
\begin{center}
{\Large \bf
On the Consistency of the Exact Renormalization Group Approach Applied to
Gauge Theories in Algebraic Non-Covariant Gauges}
\end{center}
\par \vskip 2mm
\begin{center}
{\large M.\ Simionato}
\vskip 5 mm
{\small \it
LPTHE, Universit\'e Pierre et Marie Curie (Paris VI) et Denis Diderot 
(Paris VII), Tour 16, $1^{er}$ 
\'etage, 4, Place Jussieu, 75252 Paris, Cedex 05, 
France and Istituto Nazionale di Fisica Nucleare (INFN) Frascati, Italy\\
E-mail: micheles@lpthe.jussieu.fr}
\end{center}
\par \vskip 2mm
\begin{center} {\large \bf Abstract} 
\end{center}
{\small \begin{quote}
We study a class of Wilsonian formulations
of non-Abelian gauge theories in algebraic non-covariant gauges
where the Wilsonian infrared cutoff $\Lambda$ is inserted as a mass 
term for the propagating fields. In this way the Ward-Takahashi identities 
are preserved to all scales. Nevertheless BRST-invariance
in broken and the theory is gauge-dependent and unphysical at $\Lambda\neq0$.
Then we discuss the infrared limit $\Lambda\to0$. We show 
that the singularities of the axial gauge choice are avoided
in planar gauge and light-cone gauge. In addition 
the issue of on-shell divergences
is addressed in some explicit example. Finally the rectangular Wilson loop 
of size $2L\times 2T$ is evaluated at lowest order in perturbation theory and
a noncommutativity between the limits $\Lambda\to0$ and $T\to\infty$ is
pointed out.\\ 

\end{quote}}
\end{titlepage}

\section{Introduction}

In recent years, a lot of activity has been devoted to the study of Quantum
Field Theory in the framework of the Exact Renormalization Group 
Equation (ERGE) of Wilson \cite{Wilson}. The principal advantage of this 
approach is the fact that the ERGE can be solved numerically at the 
non-perturbative level: this feature in principle pave the way to a number of 
important applications. However, in practice,
the method works well 
for scalar theories, where there is an extensive literature and a collection 
of reliable numerical results (see \cite{Bervillier} for a recent review and
a complete bibliography) 
whereas its application to gauge theories should be regarded as problematic. 
The root of the difficulties becomes clear if we remind the logic 
of the ERGE approach in the formalism of the one-particle 
irriducible (1PI) effective action 
\cite{Wetterich1,BDM}, which is the following: i) take a given Quantum 
Field Theory and introduce a fictitious infrared cutoff $\Lambda$ by modifing
the bare propagator in some way;
ii) write the functional identity which describes how the 1PI effective
action of the theory changes when the infrared cutoff is lowered;
iii) fix the boundary conditions at the ultraviolet scale $\Lambda_0$ by
identifying the effective action with the local bare action 
and integrate down numerically \cite{Wetterich1} or analytically in
perturbation theory \cite{BDM} the equation up to the the infrared value 
$\Lambda\ll\Lambda_0$.

This logic is perfectly legitimate in scalar theories, but quite 
delicate in the case of gauge theories since there is no way of 
inserting the infrared cutoff in a way consistent with gauge-invariance
(we mean with BRST-invariance): as a consequence the $\Lambda\neq0$ theory
is gauge-dependent and
unphysical and the limit $\Lambda\to0$ must be taken strictly.
This generates two distinct problems.

The first problem, which has been extensively studied in the literature
\cite{Ellw,fine-tuning}, is the following. Since the
infrared cutoff breaks BRST-invariance, the usual
Slavnov-Taylor identities have to be modified and assume a very complicate 
form \cite{Ellw} such that it is practically impossible to 
find truncations of the effective
action consistent with the modified identities. As a consequence at
the non-perturbative level
there is no rigorous control on the fact that the gauge-invariant theory
is recovered when the infrared cutoff is removed.
On the contrary at the perturbative level this is guaranteed by a
theorem: no matter how the infrared cutoff is inserted, the $\Lambda\to0$
theory is gauge invariant
provided that we choose the ultraviolet bare action in a very precise
way, which can be established in perturbation theory to all orders by
solving the fine-tuning equations discussed in \cite{fine-tuning}.
Therefore, at least in principle we could consider solved
this problem, even if in practice the fine-tuning equations
are quite involved and cannot be solved out of perturbation theory.

The second problem, which in our knowledge has been treated only
partially in the literature,
is the question of the regularity of the $\Lambda\to0$ limit. 
Actually the smoothness of this limit is guaranteed only for the proper
vertices at non-exceptional configuration of momenta \cite{BDM.IR}: 
but these are
{\it not physical quantities}. On the contrary, the limit is extremely 
delicate in the computation of physical quantities as for instance
the interquark potential: we  will see in this paper that the
limit must be taken with great care in order to have sensible results. 
In such a situation, a naive numerical analysis of the problem is 
certainly questionable. Apparently, this subtle point has never been 
recognized in the previous literature and is explicitly analyzed here
for the first time.

The guideline of our research program consists in setting up a formalism 
where these problems can be faced analytically with perturbative 
computations, in such a way to have definite answers on the
recovering of the gauge invariance and on the
regularity or the singularity of the $\Lambda\to0$ limit for physical
quantities.
In this logic, in a couple of recent papers \cite{paper.I,paper.II} 
we presented a Wilsonian formulation of gauge theories in which
the Wilsonian infrared cutoff $\Lambda$ is introduced
as a mass-like term for the propagating fields. 
In this way the Wilson's Renormalization Group Equation for the
1PI effective action becomes consistent with
the Ward-Takahashi identities to all scales. This important property, which
allows to solve the problem of finding gauge-consistent truncations but not
the question on the regularity of the $\Lambda\to0$ limit, was
proved in the Abelian case in covariant gauges \cite{paper.I}
and in the non-Abelian case in algebraic non-covariant gauges 
\cite{paper.II}. In this latter case, the crucial point was the
fact that in algebraic non-covariant gauges the ghost fields 
decouple and the gauge-symmetry can be implemented via simple 
Ward-Takahashi identities instead that in terms
of complicated Slavnov-Taylor identities. 
However, as we said in 
\cite{paper.II}, even if the implementation of the gauge symmetry is quite
efficient in these formulations there are various 
disadvantages to be taken in account.
The first disadvantage, common also to the covariant version of the
method, is the need for an explicit regularization
of the evolution equation, since when the mass cutoff
is employed the ultraviolet momenta are not sufficiently suppressed. 
However, in our view this is a
little price to pay since, at least in perturbation theory, there exist
gauge-invariant techniques to manage the ultraviolet problem (dimensional
regularization, higher derivative regularization, etc.).
The second disadvantage is the gauge-dependence of the would-be
physical quantities at $\Lambda\neq0$. However, even this problem is
not so dramatic since the gauge-dependence
can be controled throught the generalized functional identities described
in \cite{paper.II}.
It is only the third disadvantage, i.e. the fact that  
non-covariant gauges have a very delicate $\Lambda\to0$ limit, which
is a very serious one. 
This feature gives rise to difficulties since if the $\Lambda\to0$
limit does not exist for quantities which are instead well
defined in other gauges the whole approach should be considered as
suspicious. In practice this fact
does not prevent phenomenological applications since there 
are infrared safe quantities which can be computed in some approximation 
(i.e. by a truncation of the evolution equation) without encountering 
problems. 
Nevertheless in questions of principle 
one would like to have a formalism with
a well definite perturbative expansion and fully consistent with other 
gauges for {\it any} physical quantity. This is why in this paper 
we carefully study the infrared limit of the non-Abelian theory in different
algebraic non-covariant gauges. We will see in particular that the 
planar gauge and the light-cone gauge are regular in the $\Lambda\to0$
limit whereas the axial gauge is singular.
The origin of the problems of the axial gauge will be connected
with the spurious infrared divergences introduced by the double pole term 
in the propagator, i.e. the term proportional to 
$n^2q_\mu q_\nu/[n\cdot q]^2$ .
This is hardly unexpected, since it is well known that in the standard approach
using the Cauchy Principal Value (CPV) prescription the singularities
connected with the double pole are responsible for many serious problems, as
the loss of unitarity \cite{Naka} and the failure of the Wilson loop 
consistency check \cite{Soldati}. In particular this latter quantity is
the most interesting to study since it is connected with the interquark
potential, which is the typical quantity one would like to compute 
non-perturbatively in the Wilson renormalization group approach
\cite{Ellw.pot}. We recall that problems are expected to be present in the
perturbative computation of this quantity since it was shown long time 
ago, working in the temporal
axial gauge \cite{CCM} that the Wilson loop of sides $2L\times 2T$ which
directly corresponds to the interquark
potential computed at order $g^4$ is 
{\it different} from the corresponding computation in covariant gauges. 
More dramatically, in the standard perturbative expansion, 
the temporal gauge Wilson loop {\it does not} 
exponentiate in the limit $T\to\infty$ i.e. it is
{\it not} of the form
\formula{esp.}
{W_{\Gamma_{LT}}\equal{T\to\infty}\exp[-i\;2T\;V(2L)]\;,}
(see \cite{GC.Rossi} for a critical analysis).
The same
problem was found in the spatial axial gauge  for
loops in the Euclidean space with the side $T$ in the direction of the
axial vector $n^\mu$
\footnote{Obviously there
exist loops which give the same result in axial gauge and in 
Feynman gauge and in general smooth loops are expected to be safe.}
\cite{Soldati}.
The origin of the problem was found in the  Cauchy Principal Value 
Prescription which in fact breaks gauge-invariance. 
There has been a lot of work in the literature to try to solve this
problem (see the references cited in the monograph 
\cite{Bassetto}).
In the Wilsonian approach one naturally obtains a regularization 
of spurious divergences different from the CPV prescription and which 
is consistent with unitarity. Nevertheless, as we anticipated in 
\cite{paper.II}, still one expects to have problems in the infrared limit 
$\Lambda\to0$. 
These problems are discussed in detail in this paper: in addition we prove
that they are {\it absent} in more stable algebraic non-covariant gauges,
as  the planar gauge and the light-cone gauge. This is not a surprise since
it is well known in the literature that these gauge are safe with 
traditional prescriptions (the CPV for the planar gauge and the
Mandelstam-Leibbrandt (ML) for the light-cone gauge) and in particular 
the  Wilson loop consistency check is successfully passed 
\cite{Bas.Sold,Korch} at least at order
$O(g^4)$ in perturbation theory. The light-cone 
gauge in particular should be considered as a
promising starting point for our analysis. However we stress that to
prove that the Wilsonian i.e. massive formulation tends to the standard
massless formulation (which is safe) in the $\Lambda\to0$ limit is a 
subtle point since
there are infrared sensitive quantities where the massive version 
and the massless version of the same gauge choice are different, i.e.
the $\Lambda\to0$ limit is singular. The non-trivial point is to
show that this does not happen for physical quantities. In principle
this could also
be false in a finite order perturbative computation. As we said, 
one of the motivation
of this paper it to set up a workable formalism where such a
question can be answered with a reasonable analytic effort.

In order to compare and contrast with the previous Wilsonian 
literature, we remind
that a Wilsonian approach to algebraic non-covariant gauges has already
been presented in \cite{Litim,Geiger}; nevertheless our 
analysis differs from that approach, where a generic cutoff function is 
employed thus totally obscuring a number of important properties and 
difficulties of the formalism:
in particular the gauge-invariance issue and the infrared problems
can be clearly studied only within our framework. In spite of the
peculiarities of our approach, we stress that our results should be regarded 
as quite general, since there is a large class of cutoff functions 
which in the infrared behave as a mass term: for instance the most 
used cutoff in the numerical ERGE literature
\cite{Wetterich1} the exponential cutoff
\formula{exp.cut}
{R_\Lambda(q)=\frac{q^2\exp(-q^2/\Lambda^2)}{1-\exp(-q^2/\Lambda^2)}
}
is in this class since $R_\Lambda(q)\simeq\Lambda^2$ for $q^2\ll\Lambda^2$
and thus soft momenta  are effectively screened as if there were a 
mass $\Lambda^2$. This is not, however, the general case,
since for instance the sharp cutoff is not in this class\footnote{
This can be seen for instance by using the approximation
to the sharp cutoff $$R_\Lambda(q)=\frac{q^2\exp(-(q^2/\Lambda^2)^b)}
{1-\exp(-(q^2/\Lambda^2)^b)}$$ with $b\to\infty$.
In this case the infrared modes are
very strongly suppressed since $R_\Lambda(q)$ is divergent as
$q^2\to0$. The same is obtained with other regularizations of the
sharp cutoff.}. In general we expect that the situation of the
infrared limit for other cutoff functions
is worse than in the mass-cutoff case, since the gauge-symmetry is
completely broken, Ward-Takahashi identities have a cumbersome
form, and possible gauge-cancellations between different divergent
contributions are lost.
Actually one could speculate that at the
non-perturbative level, due to infinite resummations of classes of
Feynman diagram the infrared problem could be eventually solved. 
However, as a matter of fact, the perturbative expansion is the only 
controled analytical tool we have at our disposition. This is why we are
interested in giving a Wilsonian formulation in which i) a well defined
perturbative expansion can be defined and ii) it is possible to perform
practical, analytical computations with an effort not too much bigger 
than in the standard massless approach. 
In particular in this work we discuss a general method and a series
of tricks to analytically compute finite parts of
one-loop Feynman diagrams in the
$\Lambda\to0$ limit. This is relevant in preparation to a computation
of the Wilson loop up to order $O(g^4)$.

The plan of the paper is the following:
in section 2 we review the basics of the standard
gauge-fixing procedure, pointing out some subtle aspects
which are usually not noticed since an infrared regularization
consistent with BRST symmetry is assumed, which is not our case.
In section 3 we study the properties of the general linear gauge 
$L^\mu(p) A_\mu^a(p)=0$ in presence
of a mass-like cutoff. In particular the massive
axial gauge, the massive planar gauge and the massive 
light-cone gauge are analyzed.
In section 4 we discuss the problems of the axial gauge as
connected to the presence of the double pole in the propagator,
whereas in sections 5 and 6 we show that these problems are absent in planar
and light-cone gauge. In section 7 the Wilson loop
for a finite rectangular path is explicitly computed up to order $O(g^2)$
and it is shown that the physical limit $\Lambda\to0$ exists and is independent
of the gauge-fixing choice.
Section 8 resume our work and contains our conclusions.
Two appendices close the paper: appendix A is a simple but instructive exercice
that shows the gauge-dependence problem in the computation of 
a physical quantity, the pressure of a free photon gas; appendix
B collects the technical tools and tricks which are used in the text
to compute one-loop Feynman diagrams.

\section{Remarks on the gauge-fixing procedure}

In a strictly gauge-invariant theory one cannot define a
gauge-field propagator since strict gauge-invariance implies that the free
two-point function is transverse and therefore cannot be inverted. 
As a consequence the starting point of the perturbative analysis 
in terms of Feynman diagrams is missing. 
Thus, in order to define a perturbative
quantum field theory from a classical gauge invariant field theory one 
is forced to break gauge invariance through the addition of a gauge-fixing
term to the classical action\footnote{Outside perturbation theory, for
instance in the lattice approach, one can define the theory without an 
explicit gauge-fixing. However in order to show that the continuum limit 
exists, which has been rigorously proved only in perturbation theory, a 
gauge-fixing term is needed even in the lattice formulation. In general
the gauge-fixing term is unavoidable to make contact with
the perturbative formulation, which is the only one we analytically control.}. 
The way gauge-invariance is
broken depends on the chosen gauge-fixing term, which is at large
extent arbitrary: the essential requirement is that the gauge-fixing term,
depending on the gauge field and other fields 
(ghosts, antighosts and auxiliary fields), must be a BRST-cocycle:
in this hypothesis it is possible to prove that the physical
quantities are gauge-fixing independent and therefore that there are not
physical arbitrarities.

In the path integral approach the gauge-fixing is usually introduced 
by means of the Faddeev-Popov argument: since there is an infinite number of 
gauge-equivalent 
configurations giving the same contribution to
the functional integral the partition function
is ill defined. Then the problem is solved 
by integrating only on representative of any gauge
orbit. The representative are fixed by means of a gauge fixing condition
of kind $\F^a(A)=0$;
typical choices are $\F^a(A)=\partial^\mu A_\mu^a$ (covariant gauges)
or $\F^a(A)=n^\mu A_\mu^a$  (algebraic non-covariant gauges). These latter
are interesting even because they avoid the problem of Gribov copies, i.e.
the choice of the representative of the gauge orbit is unique \cite{Bassetto}.

Now it is convenient to introduce a bit of notations  (the full list of our 
notations is reported in \cite{paper.II}). 
A gauge transformations of parameters $\delta\omega=\delta\omega\cdot T=
\delta\omega^a T_a$ is denoted by
\formula{gt}
{\delta_G A_\mu= D_\mu \delta\omega,\quad \delta_G \psi=
ig\delta\omega\psi,\quad 
\delta_G \bar\psi=-ig \bar\psi \delta\omega,}
where $T_a$ are the generators of $SU(N_c)$ in the fundamental 
representation and $N_c=3$ is the number of colors. The set of physical 
fields of theory is denoted by $\phi=(A_\mu^a(p),\psi_i(p),\bar\psi^i(p))$ and
the bare action with $S_B(\phi)$; the expression
$\dd{\F}{\omega}$ denotes the differential operator
$\dd\F{A_\mu} D_\mu$ . A gauge-invariant ultraviolet 
regularization, which exists in non-chiral theories, is understood. 
With these notations the gauge-fixed partition function can be
written as
\formula{Z.FP}
{Z_\F=\int[d\phi]\det\dd{\F}{\omega}\; 
\delta(\F(A))e^{iS_B(\phi)}.
}
The usefulness of the gauge-fixed approach is based in the following 
fundamental property: if the observable $O(\phi)$ is gauge-invariant,
then its mean value in the gauge $\F^a(A)=0$, defined as
\formula{mean.value}
{<O(\phi)>_\F=
\frac1{Z_\F}\int[d\phi]\det \dd{\F}{\omega}\; 
\delta(\F(A))e^{iS_B(\phi)}O(\phi),
}
is {\em independent} of the gauge-fixing function $\F^a$:
\formula{g.i}
{\delta_G O(\phi)=0\quad\Rightarrow\quad<O>_\F=<O>_{\F'},\quad \F\neq\F'.
}
This means that in the computation of physical quantities
we have the freedom of choosing any possible gauge-fixing.

We think that it is useful to review here the {\it formal}
argument that support 
this statement with the scope of showing
where and why this argument {\it fails} in the Wilsonian formalism.

The first step consists in introducing anticommuting fields $C$ and $\bar C$ 
and commuting auxiliary fields $\lambda$ in
order to rewrite the determinant $\det \dd{\F}{\omega} $ as
a functional integral over Grassmannian variables\footnote{Here we use 
the hermiticity conditions $C=C^\dag$ and  
$\bar C=-\bar C^\dag$ \cite{Kugo} such as the relation
$(\bar C\cdot \dd{\F}{\omega} C)^\dag=\bar C\cdot 
\dd{\F}{\omega} C$ holds.
}
\formula{intro.C}
{\det \dd{\F}{\omega}\propto\int[dC d\bar C]\exp\left[-i\bar C\cdot
 \dd{\F}{\omega} C\right]
}
and the functional delta function as
\formula{intro.B}
{\delta(\F(A))=\int[d\lambda]\exp\left[i \lambda\cdot \F(A)\right]\;;
}
then the partition function (defined up to a multiplicative constant)
can be rewritten in the form
\formula{Z.BRST}
{Z_\F=\int[dCd\bar Cd\lambda d\phi]\exp\left[iS_\F
(\phi,C,\bar C,\lambda)\right]}
where the total bare action 
\formula{S.BRST}
{S_\F=S_B+\lambda\cdot\F(A)-\bar C\cdot\dede\F{A_\mu} D_\mu C
}
is invariant under the BRST-tranformation
\formula{BRST.cov}
{s A_\mu= D_\mu C,\quad s C=-\frac12 g C\times C,\quad s\bar C=\lambda,
\quad s\lambda=0
}
for {\it any} choice of the gauge-fixing function.
The cohomological property of the BRST tranformation $s^2=0$ allows us
to prove the
independence of BRST-invariant quantities from the gauge-fixing choice. 
Consider for instance a class of gauge-fixing functions $\F^a_\ep(A)$
determined by one or more continuous parameters $\ep$: then we have to
prove that physical quantities are independent of $\ep$.
This can be easily shown by observing that 
the gauge-fixing term can be rewritten as a trivial cocicle
\formulaX
{\lambda\cdot \F_\ep(A)-\bar C\cdot \dd {\F_\ep}{A_\mu}
D_\mu C=s\left(\bar C\cdot \F_\ep(A)\right).
}
Using the fact that trivial cocycles do not
contribute to the partition function, i.e.
\formula{trivial.c}
{\int[d\Phi]e^{iS_\F(\Phi)}\; s\; f(\Phi)\equiv0,  
}
(we have denoted by $\Phi=(\phi,C,\bar C,\lambda)$ the full set of fields
of the theory) one sees that the partition function is $\ep-$independent
\formulaX
{\dedi{}{\ep}Z_\ep=\int[d\Phi]e^{iS_\F(\Phi)}
s\left(\bar C\cdot \dedi{}{\ep}\F_\ep(A)\right)
=0}
and the same holds for BRST-invariant observables $O(\Phi)$ such as $sO=0$:
\formulaX
{\dedi{}{\ep}<O>_\ep=\frac1{Z_\ep}\int[d\Phi]e^{iS_\F(\Phi)}
s\left(O(\Phi)\;\bar C\cdot \dedi{}{\ep}\F_\ep(A)\right)=0.}
Unfortunately this proof is purely {\it formal}: a rigorous
analysis must face the question of infrared divergences which can invalidate 
the argument. In fact, the perturbative expansion has no meaning at all,
if we do not specify carefully as the infrared divergences are managed.
The need for an infrared regularization can conflict with BRST-invariance
and in general to show that gauge-invariance is recovered 
when the infrared regulator is removed is non-trivial since
the existence of the infrared limit for physical quantities is 
a {\em very delicate} point. To our knowledge this issue has been not
enough emphasized in the literature on the Wilson Renormalization Group, 
in spite of the fact that
it is a very crucial one. Actually, even if the recovering of the gauge
symmetry at $\Lambda=0$ has been proved for proper vertices at non-exceptional
configurations of momenta \cite{BDM.IR}, there are no theorems guaranteeing the
infrared safety of important physical quantities as the interquark potential,
which is the first thing one would like to compute in the Wilson
renormalization group approach \cite{Ellw.pot}.
In particular, as we said in the introduction, 
there are problems in its perturbative computation. 
The reason is that the fundamental property of the
exponentiation of the Wilson loop  at $T\to\infty$
is lost at $\Lambda\neq0$ since it is related to highly
non-trivial gauge-cancellations between different contributions which are
lost when BRST-invariance is broken by the infrared
cutoff. In other terms we expect a non-commutativity of limits 
$T\to\infty$ and $\Lambda\to0$. This can be seen even at $O(g^2)$ in
perturbation theory using the non-covariant formalism and it will
be discussed at length in section 7. This peculiarity suggests 
that numerical computations of the interquark 
potential should be performed with great care.

\section{The general linear gauge}

In general we are interested in {\it linear} gauges where 
$\F(A)$ is a linear function of the gauge field,
\formula{lin.gauge}
{\F(A)=L^\mu(p) A_\mu(p),\quad L_\mu(p)=a n_\mu
+b p_\mu.}
Notice that this is a class of interpolating gauges involving both
the covariant gauges (if $a=0$) and the algebraic non-covariant
gauges (if $b=0$). Other non-covariant
gauges, as for instance the Coulomb one, can be obtained for
particular values of $a$ and $b$.
In order to integrate out the auxiliary fields 
it is convenient to introduce a small convergence factor in the
functional integral measure by adding to the BRST action a term
$$
\int_x\frac1{2}{\xi_2} \lambda\cdot \lambda,
$$
where $\xi_2$ is a positive parameter of mass dimension $-2$.
This terms is a trivial cocycle and therefore does not change the essential
properties of the system; in particular 
by using standard techniques \cite{Piguet}
it is possibile to prove that physical quantities are formally
independent of $\xi_2$.
By using the linear equations of motion 
\formula{B.eq}
{\lambda^a=-\frac1{\xi_2}L^\mu A^a_\mu}
we can eliminate
the auxiliary fields $\lambda^a$ and consider the reduced action
\formula{S.linear}
{S^{red}_{lin.g.}=\int_x-\frac14F_{\mu\nu}\cdot F^{\mu\nu}
-\bar C\cdot L^\mu D_\mu C+
\int_x\frac1{2\xi_2}L^\mu A_\mu\cdot L^\nu A_\nu\;.
}
To this reduced three-level action, we add an infrared cutoff
as a mass-like term \cite{paper.II}
\formula{Gamma.0.W}
{S_{lin.g.}^{red}(A,C,\bar C,\Lambda)=S^{red}_{lin.g.}(A,C,\bar C)
+\int_x\frac12\Lambda^2A^2.
}
Now we can define an invertible massive propagator
for the gauge fields.
The explicit form of the propagator can be obtained in 
Euclidean space (Euclidean notations are recalled in appendix B)
by inverting the matrix
\formula{interp.D}
{
D^{-1}_{\Lambda,\mu\nu}(p_E)=
\left(t_{\mu\nu}(p_E)p_E^2+\frac1{\xi_2}L_\mu(p_E) L_\nu(p_E)+\Lambda^2
\delta_{\mu\nu}\right),
}
where $t_{\mu\nu}(p_E)$ denotes the transverse projector in Euclidean space
\formula{def.t} 
{t_{\mu\nu}(p_E)=\delta_{\mu\nu}-
\frac{p_{E,\mu} p_{E,\nu}}{p_E^2}.}
In particular we are interested in the $\xi_2\to0$ limit,
when one eigenvalue of the propagator vanishes, due to the transversality
property
\formula{trans}
{L^\mu D_{\Lambda,\mu\nu}(p_E)\buildrel{\xi_2\to 0}\over=0,\quad\forall
\ \Lambda
}
and therefore
$D_{\Lambda,\mu\nu}$ is not invertible (for any $\Lambda$). The final result is
\formulona{D.linear}
{D_{\Lambda,\mu\nu}(p_E)
\buildrel{\xi_2 \to 0}\over=\frac1{ p_E^2+\Lambda^2}\delta_{\mu\nu}-
\frac{p\cdot L(L_\mu p_{E,\nu}+L_\nu
p_{E,\mu})}{(p_E^2+\Lambda^2)((p_E\cdot L)^2+L^2\Lambda^2)}+\\
\frac{p_{E,\mu} p_{E,\nu}\ L^2}{(p_E^2+\Lambda^2)((p_E\cdot L)^2+
L^2\Lambda^2)}-\frac{\Lambda^2 L_\mu L_\nu}
{(p_E^2+\Lambda^2)((p_E\cdot L)^2+L^2\Lambda^2)}.
}
Notice that in limit $\xi_2\to0$
the propagator is invariant up to rescaling of the
gauge-fixing $L(p_E)\to C(p_E^2) L(p_E)$.
In particular for $L_\mu\propto n_\mu$ we recover the massive axial gauge
introduced in \cite{paper.II} which satisfies simple Ward identities,
whereas for $L_\mu\propto p_\mu$ we obtain
a massive version of the usual Landau gauge, which does not
satisfies linearly broken Ward identities, but instead satisfies 
a little modification of the usual 
Slavnov-Taylor identities (this is the Curci-Ferrari model
in Landau gauge \cite{Curci-Ferrari}).
We have also computed the explicit form of the propagator at $\xi_2\neq0$,
but the resulting expression is not particularly illuminating and
there is no scope in writing it here. 

\subsection{Planar gauge}

The  standard massless version of the
planar gauge is formally obtained from the generalized
axial gauge if we replace the parameter $\xi_2$ with the momentum
dependent function $\xi_2(p)=n^2/p^2$; in this case the BRST
action reads
\formula{S.planar.BRST}
{S^{BRST}_{planar}=
\int_x-\frac14F_{\mu\nu}\cdot F^{\mu\nu}-\bar C\cdot n^\mu D_\mu C+
\lambda\cdot n^\mu A_\mu+\frac12 \lambda\cdot\frac{n^2}{\partial^2}\lambda
}
or, after elimination of the auxiliary fields,
\formula{S.planar}
{S^{red}_{planar}=
\int_x-\frac14F_{\mu\nu}\cdot F^{\mu\nu}-\bar C\cdot n^\mu D_\mu C-
\frac12 n^\mu A_\mu\cdot\partial^2 n^\nu A_\nu.
}
The planar gauge is less problematic than the axial gauge, but the
price to pay is in a more complicate realization of symmetries:
in opposition to the axial gauge case, 
the action \rif{S.planar} does not satisfies simple
Ward-Takahashi identities (the reason being the presence of derivatives in the 
gauge-fixing term) but instead Slavnov-Taylor-like identities; moreover
ghost fields play a non-trivial role even if less crucial than in 
covariant gauges\footnote{For instance the planar gauge
has the interesting feature that all diagrams involving
ghost loops identically vanishes \cite{Bassetto,Leibbrandt}. 
This property is preserved
even in the massive version.
Nevertheless, there are non-vanishing vertex
corrections including ghosts as external lines. In principle this could
be avoided by considering a formulation without ghost fields 
\cite{Kummer} but then the Ward identities becomes quite 
cumbersome.}.
The planar gauge has been
studied quite intensively in the literature and has some interest 
in itself, and also in comparison with the axial gauge and the light-cone
gauge in the problem of inconsistencies of perturbation theory.
Moreover the computation of Feynman diagrams in planar gauge is similar
but simpler than in light-cone gauge.
For these reasons we will analyze in detail this gauge choice. We found
that the more convenient way to insert the infrared cutoff in order to have
a simple propagator is to modify the massless BRST action as
\formulona{S.planar.Lambda}
{S_{planar}(\Lambda)=&\int_x-\frac14F_{\mu\nu}\cdot F^{\mu\nu}
-\bar C\cdot n^\mu D_\mu C+\lambda\cdot n^\mu A_\mu
\\
&+\int_x\frac12\Lambda^2 A_\mu\cdot A^\mu+\frac12
\lambda\cdot\frac{n^2}{\partial^2+\Lambda^2}\lambda.
}
This gives as tree level reduced action
\formulona{Gamma.planar}
{S^{red}_{planar}(\Lambda)=&
\int_x-\frac14F_{\mu\nu}\cdot F^{\mu\nu}+\frac12\Lambda^2
A\cdot A+\\
&-\frac1{2n^2}n_\mu A^\mu\cdot(\partial^2+\Lambda^2)n_\mu A^\mu-
\bar C\cdot n^\mu D_\mu C\;.
}
Notice that the ``mass'' $\Lambda^2$ multiplies the term
$$
\frac12 A^\mu \left(g_{\mu\nu}-\frac{n_\mu n_\nu}{n^2}\right) A^\nu=
\frac12 A^ig_{ij}A^j,\quad i,j\in\{0,1,2\}
$$
and thus only transverse (with respect to $n_\mu$) degrees of freedom
are screened. There are other possible ways of introducing the infrared
cutoff in the planar gauge, but this is 
the more interesting one in the sense that
one obtains a propagator which is simple and very similar to the propagator
of the light-cone gauge.
Its explicit form in Minkowsky space is
\formula{D.planar}
{-D_{\Lambda,\mu\nu}(p)=\frac1{ p^2-\Lambda^2+i\ep}\left\{
g_{\mu\nu}-
\frac{n_\mu p_\nu+n_\nu
p_\mu}{[p\cdot n]}+\frac{\Lambda^2 n_\mu n_\nu}
{[p\cdot n]^2}\right\}.
}
The most important difference with respect to
the massive axial gauge case is the fact
that in the massive planar gauge there are {\it spurious divergences}
at $p\cdot n=0$ even at $\Lambda\neq0$. Therefore we need an explicit
prescription to manage them. The simplest choice which works, 
at least up to the order $O(g^4)$ in the stardard massless case\footnote{
Actually in higher order computations the CPV prescription could be 
problematic; see the discussion in section 5.1.}, is the CPV
prescription
\formula{CPV}
{\frac1{[p\cdot n]}\equiv\lim_{\ep\to0}\frac{p\cdot n}
{(p\cdot n)^2+\ep^2}\;.
}
We will use this prescription even in the massive case.
We stress that if we split the propagator in two pieces as
\formula{D.split}
{D_{\Lambda,\mu\nu}=\bar D_{\Lambda,\mu\nu}+\tilde D_{\Lambda,\mu\nu}
}
where $\tilde D_{\Lambda,\mu\nu}$ is the term proportional to 
$n_\mu n_\nu/n^2$,
\formula{tilde.D}
{\tilde D_{\Lambda,\mu\nu}(p)=
-\frac{\Lambda^2n_\mu n_\nu}{[p\cdot n]^2(p^2-\Lambda^2+i\ep)},
}
we can see that our formulation reduces
to the usual planar gauge expression in the 
$\Lambda\to0$ limit {\it if} the $\tilde D_{\Lambda,\mu\nu}$ term can be
neglected. This is obvious in the computation of infrared finite Feynman
diagrams, far from being obvious in the computation of infrared divergent
Feynman diagrams, and definitely non-trivial for physical quantities 
sensitive to the infrared, such as the interquark potential.
 
\subsection{Light-cone gauge}

The light-cone gauge is the most used and the most tested algebraic 
non-covariant gauge. At the present it passed many serious consistency checks
and should be considered at the same level of safety of the covariant gauge,
at least for what concerns the perturbative expansion 
in four dimensional gauge theories. 
In particular it has been explicitly 
proved at order $O(g^4)$ in perturbation theory that
the Wilson loop computed in the light-cone gauge gives the same result of
the Wilson loop in covariant gauge, provided that we use the
Mandelstam-Leibbrandt (ML) 
prescription to regularize the spurious singularities \cite{Korch}.

The Euclidean massive light-cone propagator can be formally
extracted from espression \rif{D.linear} for light-like vectors $L_\mu=
n_{E,\mu}$ such as $n_E^2=0$:
\formula{D.light-cone.E}
{D_{E,\mu\nu}(p_E)=\frac1{ p_E^2+\Lambda^2}\left\{
\delta_{\mu\nu}-\frac{n_{E,\mu} p_{E,\nu}+n_{E,\nu}
p_{E,\mu}}{p_E\cdot n_E}-\frac{\Lambda^2 n_{E,\mu} n_{E,\nu}}
{(p_E\cdot n_E)^2}\right\}.
}
The light-cone condition can be realized in Euclidean space if the
gauge vector $n_E=(n_1,
n_2,n_3,n_4)$ has
the form $n_E=(0,0,1,i)$ or $n_E^*=(0,0,1,-i)$, or an equivalent one
after a $SO(4)$ rotation. 
Writing
\formulaX
{p_3=p_\parallel\cos\theta_\parallel,
\quad p_4=p_\parallel\sin\theta_\parallel,\quad  
p_\parallel^2=p_E\cdot n_E\; p_E\cdot n_E^*=p_3^2+p_4^2,
}
we obtain the explicit expressions
\formulaX
{\frac1{p_E\cdot n_E}=\frac1{ip_4+p_3}=\frac{-ip_4+p_3}{p_4^2+p_3^2}=
\frac{p_E\cdot n_E^*}
{p_\parallel^2}=\frac{e^{-i\theta_\parallel}}{p_\parallel}.
}
After Wick rotation in Minkowsky space $p_0=-ip_4$
we see that this approach corresponds
to regularize the spurious poles with the ML prescription. The Minkowskian
propagator reads
\formula{D.light-cone}
{-D_{\Lambda,\mu\nu}(p)=\frac1{ p^2-\Lambda^2+i\ep}\left\{
g_{\mu\nu}-\frac{n_\mu p_\nu+n_\nu
p_\mu}{[[p\cdot n]]}+\frac{\Lambda^2 n_\mu n_\nu}
{[[p\cdot n]]^2}\right\},
}
with
\formula{ML}
{\frac1{[[p\cdot n]]}\equiv\frac{p\cdot n^*}
{(p\cdot n) (p\cdot n^*)+i\ep},\quad
n=(1,0,0,1),\quad n^*=(-1,0,0,1)
}
and reduces to the standard one at $\Lambda\to0$
{\it if} the $\tilde D_{\Lambda,\mu\nu}$ term can be
neglected.
Differently from the planar gauge,
the light-cone choice is extremely convenient since not only it avoids the
double pole problem but still mantains
the advantages of the axial gauge. i.e. the transversality property
which garantees the full decoupling of ghost fields,
\formula{transv}
{n^{\mu} D_{\mu\nu,\Lambda}(q)=0.}
Furthermore the simple Ward identities
\formula{WLC}
{\quad q^\mu D_{\Lambda,\mu\nu}(q)=
\frac{n_\nu}{[[n\cdot q]]}
}
and
\formula{DVD}
{p^\mu D^{\lambda\nu}_\Lambda(q) V_{\nu\mu\rho}(q,p,-q-p)
D^{\rho\tau}_\Lambda(q+p)=
D^{\lambda\tau}_\Lambda(q)-D^{\lambda\tau}_\Lambda(q+p)
}
hold, where
\formula{V...}
{V_{\mu\nu\rho}(p,q,r)=(q-r)_\mu g_{\nu\rho}+(p-q)_\rho g_{\mu\nu}+(r-p)_\nu 
g_{\rho\mu}
}
is the non-Abelian three-gluon vertex. By using these ingredients we
explicitly checked the transversality of the gluon propagator and in
general the 
Ward identities on the proper vertices with direct diagrammatic
considerations, for {\it any} value of $\Lambda$. We repeat that 
in planar gauge, on the contrary, the Ward identities
are much more complicated and the gluon propagator is not transverse
with respect to $n_\mu$ or $p_\mu$.

It is important to stress that: 
i) it is impossible to impose condition \rif{DVD}
in covariant gauges: this is the reason why the gauge symmetry is
unavoidably more complicate in the covariant case and ghost fields have
to be taken into account
ii) differently from planar gauge, we checked that
there are no other possibile forms of the light-cone propagator
consistent with \rif{transv},\rif{WLC} and \rif{DVD}: the transversality
constraint plus the symmetry requirement
fixes univocally the form \rif{D.light-cone} for the light-cone propagator.
In other words, equation \rif{D.light-cone} is 
the {\it unique} way of introducing
a Wilsonian infrared cutoff in a non-Abelian gauge theory consistently
with Ward identities.

\section{Problems of the pure axial gauge}

As we recalled in the introduction, the standard approach to axial gauge
with the CPV prescription has various problems and in particular
the test of exponentiation for a properly chosen Wilson loop fails. 
There has been a lot of work in the literature to solve this
problem (see for instance the citation list in \cite{Bassetto}),
but in spite of this effort in our opinion at present there are no completely 
satisfactory solutions.
For instance there is an apparently simple solution 
consisting in changing the 
prescriptions on the gluon propagator and the ghost interaction 
in such a way that the Wilson loop becomes the same as computed
in covariant gauges. This is the logic of Cheng and Tsai \cite{Cheng}
and also of the approches based on interpolating gauges, in which one
try to define the axial gauge as a deformation of a more regular gauge.
This kind of approaches has a long hystory (starting from the old work of
\cite{CCM} until very recent papers \cite{Joglekar,McCartor}) nevertheless
they are not completely satisfactory
since i) the transversality property $n^\mu D_{\mu\nu}=0$ is lost;
ii) the ghost fields are no longer decoupled; iii) the Ward identities
have no more a simple form; iv) a careful study of the infrared
singularities appearing in the limit in which the modified gauge tends
to the pure axial gauge is needed.
For such reasons these modifications are so drastic that in fact they
should be interpreted as a switch to a truly different gauge \cite{Bassetto}. 
The situation in the Wilsonian approach is different but still there
are problems which will be discussed in the following sections.
Our final conclusion will be that the axial gauge is definitively sick, at
least in perturbation theory, and alternative gauge choices should be
considered.

\subsection{The CPV regularization of the double pole}

In the Cauchy Principal Value prescription, the double pole is
defined as the derivative of the single pole,
\formula{CPV.double}
{
\frac1{[p_3^2]}=-\dede{}{p_3}\frac1{[p_3]}
}
where $1/[p_3]$ is defined as in \rif{CPV}.
This means that the double pole is regularized as
\formula{CPV.double.reg}
{\frac1{[p_3^2]}=\lim_{\Lambda\to0}\frac{p_3^2-\Lambda^2}
{(p_3^2+\Lambda^2)^2}\;.
}
We stress two things:
\begin{enumerate}
\item One could expect some problem with this regularization of the 
double pole, since we loose the property of the Euclidean
propagator of being positive definite. This can be seen by looking at
the eigenvalues of the Euclidean pure axial gauge propagator (at zero mass
and with the CPV prescription), which are
\formula{eigen.axial}
{0,\; \frac1{p_E^2},\; \frac1{p_E^2},\;\frac1{[p_3]^2}\;.
}
The first three eigenvalues are obviously non negative; the problem
is with the last one, which is {\it not positive definite}.
This means that the integral
\formula{neg.def}
{I_3(f)=\int_{p_3}\frac1{[p_3^2]}f(p_3)<0 }
can be {\it negative} even if $f(p_3)$ is a regular {\it positive}
function. For instance it is trivial to check that this happens
for $f(p_3)=1/(p_3^2+m^2)$.
Therefore $D_{E,\mu\nu}(p_E)$ is no more positive definite and clearly
this is a potential source of problems for the perturbative expansion.
 
\item The Wilsonian prescription of the double pole
\formula{W.prescr}
{\frac1{[p_3^2]_W}=\lim_{\Lambda\to0}\frac1{p_3^2+\Lambda^2}
}
is {\it different} from the CPV prescription \rif{CPV.double.reg}. 
As a matter of fact the 
eigenvectors of the Euclidean propagator are explicitly non-negative
\formula{eigen.axial.W}
{0,\; \frac1{p_E^2+\Lambda^2},\; \frac1{p_E^2+\Lambda^2},\;
\frac1{p_3^2+\Lambda^2}\;
}
and therefore
one could argue that the Wilsonian prescription is better then the CPV.
Unfortunately, the Wilsonian prescription has other problems that
we will describe in the next section.
\end{enumerate}

\subsection{The double pole problem in the Wilsonian formulation}

We found two major infrared problems in  the Wilsonian formulation 
of the axial gauge:
i) the Fourier transform of the propagator is divergent at $\Lambda\to0$;
ii) one loop Feynman diagrams which are infrared finite in covariant gauges
becomes divergent in axial gauge for {\it all} configurations of momenta.
In principle, this is not enough to prove the
inconsistency of this gauge, since we should prove that
at least one physical quantity is ill defined in the $\Lambda\to0$ limit.
Indeed one could think that due to miracolous cancellations
related to the Ward-identities (which are respected) physical quantities 
are indeed finite in the $\Lambda\to0$ limit as it happens for instance
for the Wilson loop
at order $O(g^2)$ (see section 7). However this is not garanteed in general.
Moreover, even if this were true, the singularity of the $\Lambda\to0$ 
limit would
forbid in practice any numerical application, since even extremely
small numerical breaking of the Ward-identities would result in enormous
(at the limit infinite) differences in the final result at $\Lambda\to0$.
This is a serious drawback of the axial gauge Wilsonian formulation.
We stress that, even if not noticed in those references, 
the problems we discuss here also apply to the formulations
in \cite{Litim,Geiger} where they are probably even worse due to the
wild breaking of gauge-symmetry. They were not recognized before simply
because they do not affect the computation of the one-loop
beta function, which is an universal quantity not
affected by the way the infrared cutoff in inserted \cite{paper.II} and
insensitive to the gauge fixing choice.

The origin of all problems comes from 
the part of the propagator proportional to $p_\mu p_\nu$,
\formula{D.pp}
{D_{\Lambda,\mu\nu}^{pp}=\frac{p_\mu p_\nu\ n^2}{(p^2+\Lambda^2)
((p\cdot n)^2+n^2\Lambda^2)}\;.
}
This quantity is a messy source of infrared divergencies due to the
identity (in the sense of distributions)
\formula{fund}
{\frac1{(p\cdot n)^2+\Lambda^2}\equal{\Lambda\to0}
\frac\pi\Lambda\delta(p\cdot n)
}
This means that for any regular function $f(p_3,\Lambda)$ such as $f(0,0)\neq0$
we have
\formula{fund2}
{\lim_{\Lambda\to0}\int_{p_3}\frac{f(p_3,\Lambda)}{p_3^2+\Lambda^2}=
\lim_{\Lambda\to0}
\frac{f(0,0)}{2\Lambda}=\infty\;.
}
We will see in next subsections 
the disastrous consequences of this fact on the computation
of various quantities.

\subsection{The $x-$space propagator.}

Here we show that even if the $x-$space propagator
\formula{D.axial.real}
{D_{\Lambda,\mu\nu}(x)=\int_p e^{-ip\cdot x}D_{\Lambda,\mu\nu}(p)
}
is perfectly defined for any $\Lambda\neq0$, the infrared limit 
$\Lambda\to0$ {\it does
not exist} due to a strong infrared divergence coming
from the double pole part of the propagator proportional to $p_\mu p_\nu$
(the double pole part proportional to $\Lambda^2 n_\mu n_\nu$
and the single pole part are regular as $\Lambda\to0$). 
This can be proved with a direct computation by using the result
\rif{fund2}. One obtains $D_{\Lambda,i3}=D_{\Lambda,3i}=D_{\Lambda,33}=0$ and
\formula{D.infty}
{D_{\Lambda,ij}(x)\equal{\Lambda\to0}\frac1{2\Lambda}\int_{\bar p}
\frac{p_i p_j}{\bar p^2}
e^{-i\bar p\cdot\bar x}=\frac1{2\Lambda}\frac13\delta_{ij}
\delta^{(3)}(\bar x)\equal{\Lambda\to0}\infty\;.
}
Therefore the Fourier transform of the propagator does not exist at
$\Lambda\to0$.

The level of danger of this problem is unclear at this level, 
since it could not
affect physical quantities. We will see in section 7 that the simplest
physical quantity we can compute, i.e. the Wilson loop at order $O(g^2)$, 
only depends on the Fourier transform of the {\it transverse} 
part of the propagator,
\formula{simplest}
{D^T_{\Lambda,\mu\nu}(x)\equiv\int_p e^{-ip\cdot x}\;t_{\mu\lambda}(p)
\;D_\Lambda^{\lambda\rho}(p)\;t_{\rho\nu}(p),
}
which is {\it safe} in the limit $\Lambda\to0$. To show this point 
we have to prove that the $\tilde D_{\mu\nu,\Lambda}(x)$ part of
the propagator, 
\formula{Dnn}
{\tilde D_{\Lambda,\mu\nu}(x)=D_\Lambda^{nn}(x)\frac{n_\mu n_\nu}{n^2}
}
which is the only one which
transverse part depends on the gauge vector $n_\mu$ 
is vanishing at $\Lambda\to0$.
A direct computation in Euclidean space gives
\formula{axial.lin.van}
{D_{\Lambda}^{nn}(x_E)=
\int_{p_E} \frac{\exp(i p_E\cdot x_E)}{p_E^2+\Lambda^2}\left[
-\frac{\Lambda^2}{p_3^2+\Lambda^2}\right]\;.
}
By direct inspection using \rif{fund}
one sees that this term is linearly vanishing with
$\Lambda$; therefore the infrared limit of the transverse part of the
propagator exists 
finite and is independent of the gauge vector $n_\mu$:
\formula{corr.0}
{D_{\Lambda,\mu\nu}^T(x_E)\equal{\Lambda\to0}\left(\delta_{\mu\nu}-
\frac{\partial_{E,\mu}\partial_{E,\nu}}{\partial_E^2}\right)
\frac1{4\pi^2x_E^2}.
}
The crucial point here is the fact that the singular $D_{\Lambda,\mu\nu}^{pp}$ 
term \rif{D.pp} does not contributes to the transverse part of the propagator.

\subsection{Computation of a one-loop integral}

The simplest perturbative quantity sensitive to the
double pole problem is the one-loop quark self-energy. 
In this section we will see that the $p_\mu p_\nu$ part of the propagator
gives an infinite contribution to this quantity in the $\Lambda\to0$ limit.
Unfortunately this is a gauge-dependent quantity therefore in principle
we cannot positively
conclude about the inconsistency of the axial gauge choice with the
Wilsonian prescription \rif{W.prescr}. However in practice this is a
serious drawback since it forbids
the direct computation of any perturbative quantity
except the one-loop anomalous dimensions and the beta function \cite{paper.II}.

In order to simplify the computation, by avoiding the inessential 
complications with the gamma matrices, we consider the self-energy of a {\it 
scalar} quark. Moreover, in order to avoid infrared divergences we take
a quark mass $m\neq0$ and in order to avoid ultraviolet
divergences we derive twice with respect to $m^2$. 
In this way we obtain a quantity corresponding to the four-point
vertex with two insertions at zero momentum shown in figure~1.
\figura{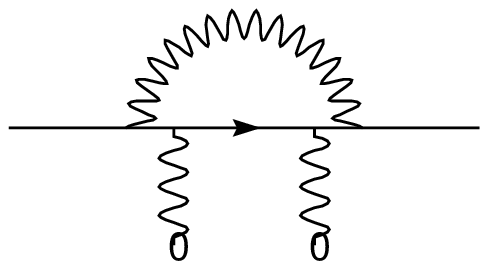}{5 cm}{Self-energy with two zero-momentum
insertions.}
This quantity, which we denote by
\formula{def.F}{
F(\bar p^2,p_3^2,\Lambda)=
\int_{q_E}\left(\dede{}{m^2}\right)^2\left[ 
\frac{(2p_E+q_E)^\mu(2p_E+q_E)^\nu}{[(q_E+p_E)^2+m^2+
\Lambda^2]}D_{\mu\nu}(q_E,\Lambda)\right]
} 
is ultraviolet convergent and (in covariant gauges) infrared finite even
at $\Lambda=0$ if $p_E^2\neq -m^2$.  
Still,  the $\Lambda\to0$ limit is singular in the massive axial gauge since 
the dominant contribution comes from the double pole part of the propagator 
which, due to equation \rif{fund2}, is infinity in the $\Lambda\to0$
limit:
\formulaX{
F(\bar p^2,p_3^2,\Lambda)\equal{\Lambda\to0}
-\frac1{2\Lambda}\int_{\vec q}
\frac2{[(\bar q+\bar p)^2+m^2]^3}
q_i q_j(2p+q)^i(2p+q)^j\to\infty.
}
Notice that this problem appears even in the {\it Abelian} theory.
Moreover it is clear that the problem becomes worse in the non-Abelian case
and at higher orders in perturbation theory.
Still, in principle it is possible that 
cancellations allows
to define unambigously physical quantities even in the $\Lambda\to0$
limit, since the various Feynman diagrams should add in such a way 
that the final result at $\Lambda\to0$ is finite and independent of
the gauge fixing, i.e. the same that in covariant gauges. However,
this is an highly non-trivial point and cannot be taken for granted
at finite order in perturbation theory.

\section{The planar gauge}

For the reasons described above, we think that it is not convenient to insist
on the axial gauge and from now on we switch to other gauge choices.
The simplest examples where the difficulties of the axial gauge can 
be circumvented (at least in the standard massless formalism and 
at order $O(g^4)$ in perturbation theory) are the planar gauge and
the light-cone gauge. 
We consider first the planar gauge.

\subsection{Remarks on the planar gauge}

Properly speaking, even the standard massless planar gauge choice
with the CPV prescription is not completely 
controled and the consistency of the perturbative
expansion in this gauge  is an open problem; nevertheless
it will be studied in this section for sake of comparison with the axial gauge and
the light-cone gauge choices.

The reason of the difficulties
can be traced back to the well known splitting formula \cite{Bassetto}
\formula{splitting}
{\frac1{[n k]}\frac1{[n(p-k)]}=\frac1{np}\left(\frac1{[nk]}+\frac1{[n(p-k)]}\right)
-\pi^2\delta(nk)\delta(np)\;.
}
Due to the presence of the delta function terms, in two-loops or higher orders 
computations involving gluon vertices there appear ill-defined
objects of kind
\formula{ill.defined}
{\frac1{[kn]}\delta(kn)\sim\frac{kn}{(kn)^2+\ep^2}\;
\frac{\ep/\pi}{(kn)^2+\ep^2}
}
which have to be studied in detail.
It is known that in the massless
computation of the $O(g^4)$ Wilson loop there is a precise way to manage these
terms and the final result is consistent with the covariant gauge 
\cite{Bas.Sold}. 
Nevertheless, the
situation at order $O(g^6)$ is unknown and in general the definition 
of two-loops Feynman diagrams is problematic. 
This is the reason why we will switch to the light-cone
gauge in the next section.
However it is interesting to see that even at the level of the planar gauge
the divergences found before in one-loop computations are absent,
essentially
because the double pole term proportional to $p_\mu p_\nu$
in \rif{D.pp} is absent. There is however a double pole in the 
$\tilde D_{\Lambda,\mu\nu}$
part of the propagator, but since this
term is multiplied by a $\Lambda^2$ factor it is espected to vanish 
in the $\Lambda\to0$ limit. This is 
indeed the case for infrared safe quantities
and it will be explicitly verified in the
examples we consider here.
We stress that this is by no means a trivial statement, since 
in the computation of particular infrared divergent
quantities the  $\tilde D_{\Lambda,\mu\nu}$ 
term in the denominator could be not
subleading with respect to the other terms and actually could also give the
{\it dominant} contribution. We will see in section 6 that this 
is indeed the case in the light-cone gauge.  Moreover we will see
in section 7 that the $\tilde D_{\Lambda,\mu\nu}$ term contributes to the
Wilson loop and is suppressed only if the $\Lambda\to0$ limit is
done properly, i.e. at finite $T$.

\subsection{The $x-$space propagator}

We prove here that the Fourier transform of the propagator is well 
defined and in the
$\Lambda\to0$ limit reduces to the Fourier transform of the standard
planar gauge propagator. This is obvious for the $\bar D_{\Lambda,\mu\nu}$ 
part of the propagator, therefore we have simply to prove that the Fourier 
transform of the $\tilde D_{\Lambda,\mu\nu}$ term 
is vanishing at $\Lambda\to0$. A direct
computation gives
\formulaX
{\lim_{\Lambda\to0}\Lambda^2\int_{\bar p,p_3}\frac{\exp(ip_3 x_3+
i\bar p\cdot \bar x)}
{[p_3]^2(p_3^2+\bar p^2+\Lambda^2)}=\lim_{\Lambda\to0}\frac{\Lambda^2}
{8\pi^2}\ln(\Lambda^2 \bar x^2)=
0.
}
This result can be obtained by first doing the $p_3$ integral using $1/[p_3^2]=
[p_3]^{-1}\partial_{p_3}$; the successive 
three-dimensional integral generates an infrared logarithimic singularity at
$\Lambda\to0$ which is killed by the $\Lambda^2$ prefactor. 
Therefore we have solved the problem with the Fourier transform.
However we stress again that in the computation of infrared sensitive
quantities this term in general could gives an essential
contribution to the final result. Furthermore, we will see in section 7
that this term is suppressed in the Wilson loop computation only
if the $\Lambda\to0$ and $T\to\infty$ limit are done in the correct order.
Therefore the relation between the massive version of the planar gauge in the
$\Lambda\to0$ limit and the standard planar gauge in general is delicate.

\subsection{Computation of a one-loop finite quantity}\label{one.loop.comp}

Now we can study what happens in the computation of the quark self-energy
with two zero-momentum insertions, i.e. the analogous of the quantity
$F$ defined in \rif{def.F}.
If we split the Euclidean planar gauge propagator in three terms (in this
section for notational simplicity we write $p$ for $p_E$)
\formulona{D.planar.E}
{D_{\mu\nu}(p,\Lambda)&=\frac1{ p^2+\Lambda^2}\left\{
\delta_{\mu\nu}-\frac{n_\mu p_\nu+n_\nu
p_\mu}{[p\cdot n]}-\frac{\Lambda^2 n_\mu n_\nu}
{[p\cdot n]^2}\right\}\\
&=D_{\mu\nu}^{(a)}(p,\Lambda)+D_{\mu\nu}^{(b)}(p,\Lambda)+
D_{\mu\nu}^{(c)}(p,\Lambda)
}
we can split the computation in three different contributions:
\formulaX
{F(\bar p^2,p_3^2,\Lambda)=F^{(a)}(p^2,\Lambda)+F^{(b)}
(\bar p^2,p_3^2,\Lambda)+
F^{(c)}(\bar p^2,p_3^2,\Lambda)}
with
\formula{F.a}
{F^{(a)}(p^2,\Lambda)=\int_{q}
\left(\frac{\partial}{\partial m^2}\right)^2\frac{(2p+q)^2}{(q^2+\Lambda^2)
((q+p)^2+m^2+\Lambda^2)},
}
\formula{F.b}
{F^{(b)}(\bar p^2,p_3^2,\Lambda)=\int_{q}
\left(\frac{\partial}{\partial m^2}\right)^2\frac{-2(2p+q)\cdot n\;(2p+q)
\cdot q}{[q\cdot n](q^2+\Lambda^2)
((q+p)^2+m^2+\Lambda^2)},
}
\formula{F.c}
{F^{(c)}(\bar p^2,p_3^2,\Lambda)=\int_{q}
\left(\frac{\partial}{\partial m^2}\right)^2\frac
{-\Lambda^2(2p\cdot n+q\cdot n)^2}{[q\cdot n]^2(q^2+\Lambda^2)
((q+p)^2+m^2+\Lambda^2)}.
}
The first contribution is the same as in Feynman gauge and
can be computed using the standard Feynman parametrization. It gives
\formula{F.a.x}
{F^{(a)}(p^2,\Lambda)=\frac1{16\pi^2}\int_0^1dx\;(1-x)^2\frac{p^2(1+4x-x^2)+2[
m^2(1-x)+\Lambda^2]}
{[(p^2x+m^2)(1-x)+\Lambda^2]^2}
}
where the $x-$integral can be performed in terms of elementary functions
(logarithms).
In particular in the on-shell limit $p^2\to-m^2,\;\Lambda\to0$ the 
$x\simeq1$ integration region dominates and we have
\formula{Fa.on-shell}
{F^{(a)}(p^2,\Lambda=0)=-\frac1{4\pi^2}\frac1{p^2+m^2}
+O((p^2+m^2)^0)
}
if the $\Lambda\to0$ limit is taken first. On the contrary, if the
 $p^2\to-m^2$ limit is taken first, we have
\formula{Fa.on-shell.2}
{F^{(a)}(p^2=-m^2,\Lambda)=-\frac1{16\pi}\frac 1{m\Lambda}
+O\left(\frac{\Lambda^0}{m^2}\right).
}
This is an explicit confirmation of the fact the the 
order of limits is a delicate question for 
infrared divergent quantities.

The second contribution is typical of planar gauge. Its general expression
is rather complicate, but there is a simplification if we restrict the
analysis to the limits $|\bar p^2|\gg p_3^2$ or $|\bar p^2|\ll p_3^2$.
The first limit $|\bar p^2|\gg p_3^2$ 
is trivial in the sense that the integral can be
recast in a covariant-like form and its computation is no more
difficult than the computation of the covariant term and gives a
completely analogous contribution. In the second
limit $|\bar p^2|\ll p_3^2$ instead the peculiarities of the planar gauge are
evident. In particular the quantity 
\formula{Fb}
{F^{(b)}(p_3^2,\Lambda)=\lim_{\bar p\to0}F^{(b)}(\bar p^2,p_3^2)
} 
can be computed with the method of double Feynman parametrization discussed in
appendix B. After some manipulation it reads
\formulaX
{F^{(b)}(p_3^2,\Lambda)=\int_{q_3,\tilde q}
\int_0^1 dx\int_{1-x}^1dz
\left(\frac{\partial}{\partial m^2}\right)^2\frac{N(q_\parallel^2,\tilde q^2,
p_3^2,x,z)}{D(q_\parallel^2,\tilde q^2,p_3^2,m^2,\Lambda^2,x,z)}
}
with
\formulaX
{N=-2[p_3^4(1-x^2)^2+p_3^2[2(3x^2-1)q_3^2-\tilde q^2(1-x^2)/z]+q_3^2[q_3^2
+\tilde q^2/z]]
}
and
\formulaX
{D=z^{3/2}[q_3^2+\tilde q^2+\Lambda^2z+(p_3^2x+m^2)(1-x)]^3.
}
Now the integrations in $q_3,\tilde q$ and $z$ are straightforward
(see appendix B) and all the complexity is confined into the $x-$integration.
In the $\Lambda\to0$ limit there is a strong simplification and it can be
exactly performed in terms of elementary functions. The final result is
\formulonaX
{F^{(b)}(p_3^2,\Lambda=0)&=-\frac1{4\pi^2p_3^2}-\frac{p_3^2/m^2-1}
{8\pi^2(p_3^2+m^2)}
\\ &-\frac{\sqrt{p_3^2}\ln[(\sqrt{p_3^2+m^2}-\sqrt{p_3^2})
/(\sqrt{p_3^2}+\sqrt{p_3^2+m^2})]}{8\pi^2(p_3^2+m^2)^{3/2}}
.}
The third contribution is characteristic of the Wilsonian formulation and
must vanish in the $\Lambda\to0$ limit in order to recover the 
results of the standard massless planar gauge:
\formulaX
{\lim_{\Lambda\to0}F^{(c)}(p_E)=0.}
This is immediate to see
in the limit $|\bar p^2|\ll p_3^2$. 
In the opposite limit $|\bar p^2|\gg p_3^2$ one
could expect divergences at $\bar p^2=-m^2$; but
it is immediate to see that
\formula{subleading}
{F^{(c)}(\bar p^2=-m^2,\Lambda)\sim\frac{\Lambda^2}{m^2}F^{(a)}(\bar p^2=
-m^2,\Lambda)
}
therefore $F^{(c)}(\bar p^2=-m^2,\Lambda)$ is linearly vanishing at
$\Lambda\to0$. Thus the $\tilde D_{\Lambda,\mu\nu}$ 
term can always be neglected, at least in this kind of one-loop
computations.
We will see that this is {\it not} always the case for the light-cone gauge.

\section{Safeness of the light-cone gauge}

For the point of view of the consistency of the perturbative expansion
the light-cone gauge is expected to be the safest choice. There are
various reasons for this expectation, which we will discuss now.

The first reason is the fact
that, contrary to the planar gauge, the Euclidean propagator
is (semi) {\it positive definite}. This can be
immediately proved by solving the eigenvalue equation
\formula{eigen.eq}
{\det\left(\lambda\delta^\mu_\nu-D_\nu^\mu(p_E,n_E;\Lambda)\right)=0
}
which after explicit computation reads
\formulaX
{\lambda^2\left(\lambda-\frac1{p_E^2+\Lambda^2}\right)^2=0
}
with solutions
\formula{eigenvalues}
{\lambda_1=\lambda_2=\frac1{p_E^2+\Lambda^2},\quad\lambda_3=\lambda_4=0.
}
From this computation one sees that the eigenvalues
$\lambda_i=\lambda_i(p_E,n_E)$ are non-negative. In addition, they 
depend only on the Lorentz-invariant
combination $p_E^2$ and not on $n_E$. We emphasize that this is {\em not true} in
non-covariant gauges  others than the light-cone one. Therefore one expects
a somehow less prononced Lorentz breaking in the light-cone gauge.

The second reason is that within the ML prescription
the splitting formula \rif{splitting} 
holds without the delta function term: as a consequence there are no ill
defined terms in higher
orders of perturbation theory 
as it happens with the CPV prescription in axial and
planar gauges.
  
The third reason is that the massless limit of the light-cone
gauge is expected to be less singular than in others gauges.
This expectation comes from the 
that the Fourier transform of the $\tilde D_{\Lambda,\mu\nu}$ term
is {\it quadratically} vanishing as $\Lambda^2\to0$: this should be
contrasted with the axial gauge case where the suppression of
the $\tilde D_{\Lambda,\mu\nu}$ term in only linear in $\Lambda$ 
(see the discussion
after Eq. \rif{axial.lin.van}) and the planar gauge where 
this term vanishes as $\Lambda^2$ times a logarithm. 
We will discuss in detail this latter point just below.

\subsection{The $x-$space propagator}

In order to explicitly compute 
the Fourier transform of the $\tilde D_{\Lambda,\mu\nu}(x)$ term
we have to study the Euclidean integral
\formula{FT.D}
{D^{nn}_\Lambda(x_E)=-\Lambda^2\int_{q_E}\frac{e^{iq_E\cdot x_E}}
{(q_E\cdot n_E)^2(q_E^2+\Lambda^2)}.
}  
We proceed as follow. First,
we introduce the angles $\theta$ and $\theta'$ such as
\formula{angles}
{q_1=q_\perp\cos\theta',
\quad q_2=q_\perp\sin\theta',\quad q_3=q_\parallel\sin\theta,\quad
q_4=q_\parallel\cos\theta
}
and we take $x=(x_\perp,0, x_\parallel,0)$ (this is not restrictive); 
then the Fourier transform can be written
\formulaX
{D^{nn}_\Lambda(x_E)=-\Lambda^2\int \frac{q_\perp dq_\perp d\theta'}
{(2\pi)^2}\frac{dq_\parallel d\theta}
{(2\pi)^2}\frac{e^{-2i\theta-iq_\parallel x_\parallel\sin\theta-
iq_\perp x_\perp\sin\theta'
}}
{q_\parallel(q_\parallel^2+q_\perp^2+\Lambda^2)}.
}
$D^{nn}_\Lambda(x_E)$
can be computed by first performing the angular integrations and then
the momentum integrations in $q_\parallel$ and $q_\perp$.
The angular integrations can
be done by using the following representation of Bessel functions
\formula{Bessel}
{J_n(z)=\frac1{2\pi}\int_0^{2\pi} d\theta\; e^{-in\theta-iz\sin\theta}
}
for $n=0$ and $n=2$ respectively. In this way we obtain
\formulaX
{D^{nn}_\Lambda(x_E)=-\frac{\Lambda^2}{(2\pi)^2}\int_0^\infty dq_\perp q_\perp
\int_0^\infty\frac{dq_\parallel}{q_\parallel}\frac{J_2(q_\parallel 
x_\parallel)J_0(q_\perp x_\perp)}{q_\parallel^2+q_\perp^2+\Lambda^2}.
} 
Since at small $z$ the Bessel function $J_2(z)\simeq\frac18z^2$ is 
quadratically vanishing 
whereas at large $z$ is
exponentially  suppressed, we see that the $q_\parallel-$integral
is infrared and ultraviolet finite. It can be exactly computed with the result
\formulaX
{D^{nn}_\Lambda(x_E)=-\frac{\Lambda^2}{8\pi^2}
\int_0^\infty dq_\perp q_\perp \frac{\omega_\perp^2-4x_\parallel^{-2}+
2\omega_\perp^2
K_2(\omega_\perp x_\parallel)}
{\omega_\perp^4}J_0(q_\perp x_\perp),}
with $\omega_\perp^2\equiv q_\perp^2+\Lambda^2$.
Using the series expansion of the Bessel functions
$$K_2(z)=\frac{4-z^2}{2z^2}+O(z^2\log z^2),\quad
J_0(z)=1-\frac14z^2+O(z^4),$$ one sees that the
$q_\perp-$integral is also finite, even in the $\Lambda\to0$ limit where it
can be computed exactly and it gives
\formula{smallL.FT}
{D^{nn}_\Lambda(x_E)=-\frac{\Lambda^2}{16\pi^2}
\left[\frac{x_E^2}{x_\parallel^2}\log\frac{x_E^2}{x_\perp^2}-1+
O(\Lambda^2 x^2_\perp,\Lambda^2 x^2_\parallel)\right]
}
where $x_E^2=x_\parallel^2+x_\perp^2$.
Thus, $\tilde D_{\Lambda,\mu\nu}(x)$ 
is quadratically vanishing in the $\Lambda\to0$ limit. We also see that
there could be problems in the limit  $x_\parallel/ x_\perp\to\infty$
which is relevant for the Wilson loop computation. This point will be
discussed in detail in section 7.

\subsection{The structure of (soft) infrared divergences}

It is possible to repeat the computation of section \ref{one.loop.comp} in
the light-cone gauge. The covariant contribution $F^{(a)}(p,\Lambda)$ 
is obviously
the same, whereas the contributions $F^{(b)}(p,\Lambda)$ and 
$F^{(c)}(p,\Lambda)$ are
different from the analogous one in planar gauge.
Still, one can use the double
Feynman parametrization method described in appendix B. In the
limit $p_\perp^2/p_\parallel^2\to0$
the contribution $F^{(b)}$ can be rewritten in the form
\formulaX
{F^{(b)}(p_\parallel^2,\Lambda)=\int_{q_\parallel,\tilde q_\perp}
\int_0^1 dx\int_{1-x}^1dz
\left(\frac{\partial}{\partial m^2}\right)^2\frac{N(q_\parallel^2,\tilde q_\perp^2,
p_\parallel^2,x,z)}{D(q_\parallel^2,\tilde q_\perp^2,p_\parallel^2,m^2,\Lambda^2,x,z)}
}
with
\formula{N.lc.x->1}
{N(q_\parallel^2,\tilde q_\perp^2,p_\parallel^2,x,z)=-8q_\parallel^2p_\parallel^2+
\Delta N(q_\parallel^2,\tilde q_\perp^2,p_\parallel^2,x,z)
}
and
\formulaX
{D=z[q_\parallel^2+\tilde q_\perp^2+\Lambda^2z+(p_\parallel^2x+m^2)(1-x)]^3.
}
Actually, we  exactly computed the  expression $\Delta N(q_\parallel^2,
\tilde q_\perp^2,
p_\parallel^2,x,z)$ in the numerator and it is possible to perform 
exactly the integrals in the general case, but the resulting expressions 
are lenghty and they will not be reported here. 
In order to analyze the on-shell divergences at $p^2\simeq-m^2$
the only relevant term is the fist one displayed in the right hand side
of \rif{N.lc.x->1}. 
The integrations in $q_\parallel$ and $q_\perp$ are straighforward
and they give, after neglecting the $\Delta N$ term, 
\formula{F2b.domin}
{F^{(b)}(p_\parallel^2,\Lambda)\simeq-\frac1{4\pi^2}\int_0^1 dx
\int_{1-x}^1\frac{dz}z\frac{p_\parallel^2(1-x)^2}
{[\Lambda^2z+(p_\parallel^2 x+m^2)(1-x)]^2}.
}
If we take the $\Lambda\to0$ limit first we obtain
\formula{F2b.1}
{F^{(b)}(p_\parallel^2,\Lambda=0)
\equal{p^2\to-m^2}-\frac1{4\pi^2}\frac{\ln[
(p_\parallel^2+m^2)/p_\parallel^2]}
{p_\parallel^2+m^2}+O\left(\frac1{p_\parallel^2+m^2}\right)
}
and this is
dominant with respect to the divergence $1/(p^2+m^2)$ coming from
the covariant contribution $F^{(a)}$. If we take the  $p^2\to-m^2$
limit first we obtain
\formula{F2b.2}
{F^{(b)}(p_\parallel^2=-m^2,\Lambda)
=\frac1{8\pi^2}\frac1{\Lambda^2}+O\left(\frac{\Lambda^0}{m^2}\right),
}
which is still dominant with respect to the linear divergence coming from
the covariant contribution $F^{(a)}$. 

Now we can study in an analogous way
the contribution of the $\tilde D_{\Lambda,\mu\nu}$
part of the propagator, i.e. the $F^{(c)}$ term. It is obvious that
this term is quadratically 
vanishing at $\Lambda\to0$ if $p^2\neq -m^2$. Nevertheless,
if  $p^2=- m^2$, this term could be divergent as $\Lambda\to0$. This
is the case indeed. Actually it is easy to understand that there is 
no divergence in the limit $|p^2_\parallel|\ll p_\perp^2$, using the
same argument as in equation \rif{subleading}.
However in the opposite limit $|p^2_\parallel|\gg p_\perp^2$ the
infrared divergence is enhanced and actually {\it dominates} with respect
to the covariant contribution. To see this point we write down the
explicit expression of $F^{(c)}$ in the near on-shell region, obtained
with the method of double Feynman parametrization. It reads
\formula{F2c.domin}
{F^{(c)}(p_\parallel^2,\Lambda)=
\int_{q_\parallel,\tilde q_\perp}
\int_0^1 dx\int_{1-x}^1dz
\left(\frac{\partial}{\partial m^2}\right)^2\frac{\tilde
N(q_\parallel^2,\tilde q_\perp^2,
p_\parallel^2,x,z)}
{\tilde D(q_\parallel^2,\tilde q_\perp^2,p_\parallel^2,m^2,\Lambda^2,x,z)}
}
with
\formula{N.leading}
{\tilde N(q_\parallel^2,\tilde q_\perp^2,p_\parallel^2,x,z)
=24(1-x)^2p_\parallel^4\Lambda^2+\Delta 
\tilde N(q_\parallel^2,\tilde q_\perp^2,p_\parallel^2,x,z)
}
and
\formulaX
{\tilde D=z[q_\parallel^2+\tilde q_\perp^2+\Lambda^2z+
(p_\parallel^2x+m^2)(1-x)]^4.
}
In the term $\Delta \tilde N(q_\parallel^2,\tilde q_\perp^2,
p_\parallel^2,x,z)$ in the numerator we collected all the contributions
which are subleading in the on-shell limit. After double derivation 
with respect to the mass $m^2$ and momentum integration we obtain
\formula{F2c.xz}
{F^{(c)}(p_\parallel^2,\Lambda\to0)\equal{p_\parallel^2\to-m^2}
\frac3{2\pi^2}\int_0^1 dx
\int_{1-x}^1\frac{dz}z\frac{\Lambda^2 p_\parallel^4(1-x)^4}
{[\Lambda^2z+(p_\parallel^2 x+m^2)(1-x)]^4}.
}
In order to study the $\Lambda\to0$ limit at $p_\parallel^2=-m^2$
it is convenient do rescale the $z$ variable as
\formula{z.res}
{z=\tilde z(1-x),\quad \int_{1-x}^1\frac{dz}z=\int_1^{1/(1-x)}\frac{d\tilde
z}{\tilde z}
}
We see that for $x\to1$ the upper limit of the $\tilde z-$integral
tends to infinity, therefore the $z-$integration simplifies. Then the
$x-$integration is done taking in account that the $x\simeq1$ integration
region dominates in the near on-shell region; finally one obtains
\formulaX
{F^{(c)}(p_\parallel^2=-m^2,\Lambda)\equal{\Lambda\to0}\frac1{6\pi^2}
\frac{m^2}{\Lambda^4}.
}
It is important to notice that even if the $\tilde D_\Lambda$ term dominates
the soft divergence at $m\neq0$, it can be always neglected in the analysis of
hard or collinear divergences, i.e. if $m=0$. In this case in fact
the contribution from the $\tilde D_\Lambda$ term is quadratically vanishing.
This point can be understood with a dimensional argument and 
explicitly checked by using the double Feynman
parametrization formula and by noticing that at $m^2=0$ it is the
$x\simeq0$ region of integration which dominates,  not the $x\simeq1$ region.
Then the $z-$integration can be done by using
\formulaX
{\int_{1-x}^1\frac{dz}zf(x,z)\simeq xf(x,1),\quad x\simeq0
}
and the $x-$integration by using the tricks reported in appendix B.
With similar techniques we checked that the gluon self-energy integral in the
$\Lambda\to0$ limit goes smoothly to the standard massless integral
for off-shell Euclidean momenta $p_\parallel^2\neq -p_\perp^2$.

\section{The Wilson loop test}

The Wilson loop is the simplest physical quantity 
where the effects and the problems of the infrared regularization can be 
studied. In particular our scope here is to study the subtilities of the
$\Lambda\to0$ limit and to test how the essential
property of the gauge-invariance of
the Wilson loop is recovered when the infrared cutoff is removed.
In concrete in this section we compute the Wilson up to order $O(g^2)$ 
in perturbation theory. 
This is enough for elucidating various important features of massive 
axial, planar and light-cone gauges and it is a first step versus a more
comprehensive computation at order $O(g^4)$ in perturbation theory.

For definiteness, we shall consider a rectangular Wilson loop 
$\Gamma_{LT}$  of size $2L\times 2T$, with 
$T\gg L$. We shall work in Euclidean space with coordinates 
$x_1,x_2, x_3, x_4$ and we shall
take the loop in the plane $x_2x_3$, as shown in figure 2. 
We should notice that $T$ denotes a
lenght in the spatial direction $x_3$, i.e. the direction of the gauge
vector $n_E^\mu=(0,0,1,0)$, and not in the temporal direction
$x_4=ix_0$. Therefore apparently this loop in not related to the interquark 
potential. Nevertheless if the theory is consistent all directions in the
Euclidean space must be physically equivalent at $\Lambda\to0$, therefore 
the final result must
be the same that for the loop corresponding to the interquark potential
where the side $T$ is in the temporal direction.

\figura{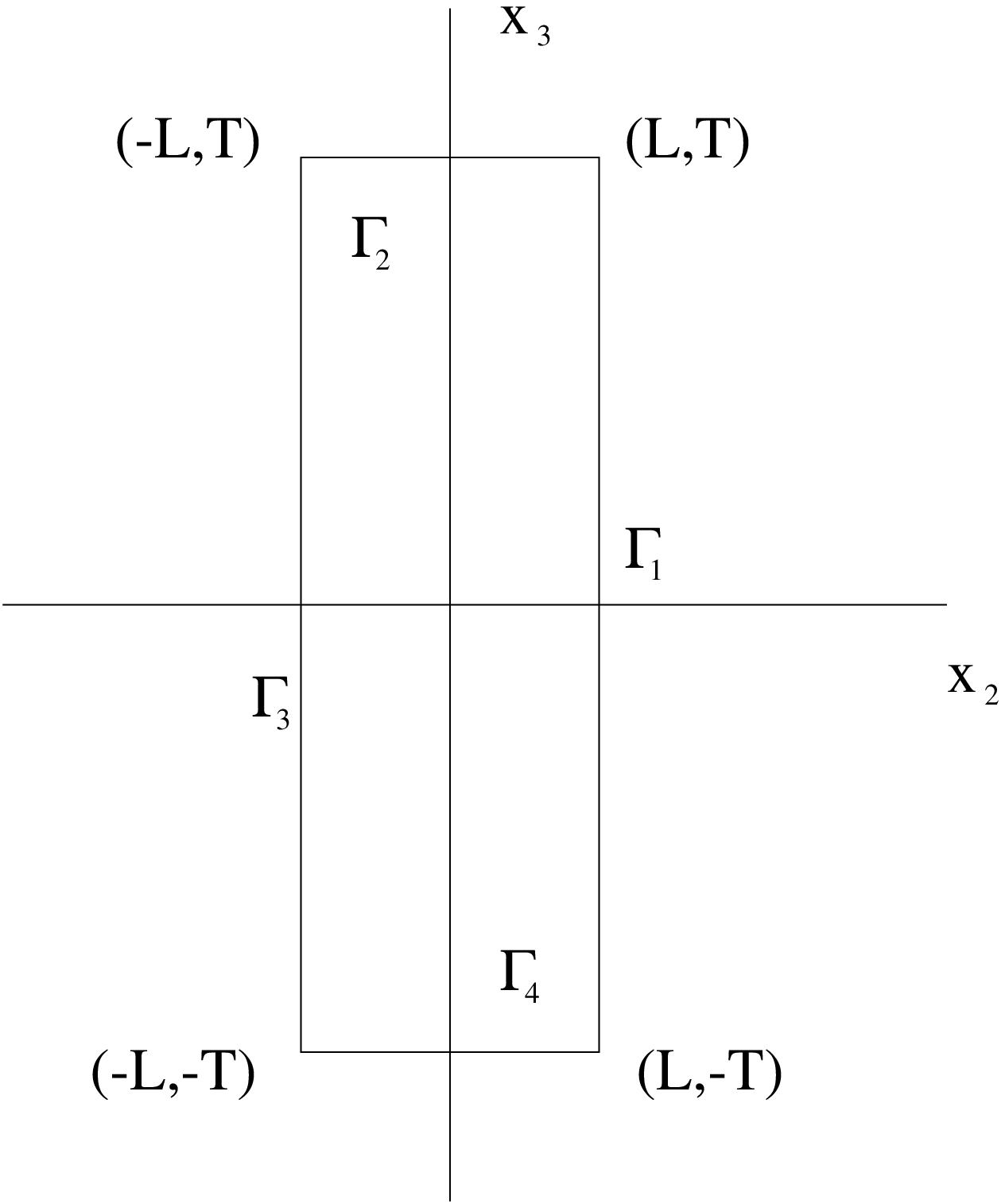}{8 cm}{Rectangular Wilson loop of size $2L\times 2T$.}

Formally the Wilson loop is defined by
\formula{rect.WL}
{W_{\Gamma_{LT}}=\frac1{N_c}
<\Tr P\exp\left(ig\int_{\Gamma_{LT}} A_\mu dx^\mu\right)>}
where $P$ denotes the path ordering on the loop
$\Gamma_{LT}$, $\Tr$ is the trace in
the fundamental representation of $SU(N_c)$ and the average is evaluated via 
a perturbative expansion of the Euclidean functional integral. 
In our case $\Gamma_{LT}$
can be split in four pieces
\formula{Gamma.LT}
{\Gamma_{LT}=\Gamma^{(1)}+\Gamma^{(2)}+\Gamma^{(3)}+\Gamma^{(4)},
}
parametrized as 
\formulonaX{
\Gamma^{(1)}(s)=sT\pmatrix{0\\ 0\\ +1\\ 0},\quad
\Gamma^{(2)}(s)=sL\pmatrix{0\\ -1\\ 0\\ 0},\\
\Gamma^{(3)}(s)=sT\pmatrix{0\\ 0\\ -1\\ 0},\quad
\Gamma^{(4)}(s)=sL\pmatrix{0\\ +1\\ 0\\ 0},
}
where the parameter $s$ lives in the interval $[-1,1]$. Equivalently,
we can parametrize 
\formula{param}
{\Gamma_\mu^{(1)}(t)=t\delta_{\mu3},\quad
\Gamma_\mu^{(2)}(l)=-l\delta_{\mu2},\quad
\Gamma_\mu^{(3)}(t)=-t\delta_{\mu3},\quad
\Gamma_\mu^{(4)}(l)=l\delta_{\mu2},
} 
with $t\in[-T,T]$ and $l\in[-L,L]$.
The Wilson loop can be easily computed by expanding in powers of the coupling
constant,
\formulaX
{W_{\Gamma_{LT}}=W_{\Gamma_{LT}}^{(0)}+gW_{\Gamma_{LT}}^{(1)}+
\frac{g^2}2W_{\Gamma_{LT}}^{(2)}+\frac{g^3}{3!}W_{\Gamma_{LT}}^{(3)}+O(g^4)\;.
}
By using the properties of the generators in the fundamental representation
of $SU(N_c)$, $\Tr\; 1=N_c$, $\Tr\; T_a=0$ and 
$\Tr(T_a T_b)=\frac12\delta_{ab}$, from \rif{rect.WL}
one obtains the explicit expression
\formula{WL.O(g^2)}
{W_{\Gamma_{LT}}=
1-\frac{g^2}{4N_c}\int_{\Gamma_{LT}}dx^\mu dy^\nu<A_\mu(x)
\cdot A_\nu(y)>^{(0)}+O(g^4).}
Notice that higher orders corrections begins at order $O(g^4)$, not $O(g^3)$.
The expectation value $<A_\mu(x)\cdot A_\nu(y)>^{(0)}$, computed at zero order
in the coupling constant, coincides with the Fourier transform of
the free propagator, therefore a a more explicit expression for the
$O(g^2)$ contribution is
\formula{WL.2}
{W_{\Gamma_{LT}}^{(2)}=-C_F\int_{-1}^1 ds_1\int_{-1}^1 
ds_2\dedi{\Gamma^\mu}{s_1}
\dedi{\Gamma^\nu}{s_2}D_{\mu\nu}(\Gamma(s_1)-\Gamma(s_2)),
}
where the relation $\delta^{aa}=2N_cC_F$ with $C_F=(N_c^2-1)/(2N_c)$ has
been used.
Specializing to the rectangular Wilson loop in figure 2 we have in
general sixteen contribution to \rif{WL.2}; however many terms gives the
same contribution due to the symmetries $D_{\mu\nu}(x)=D_{\nu\mu}(x)$ and
$D_{\mu\nu}(x)=D_{\mu\nu}(-x)$ and finally one is left with the explicit 
expression
\formulonaX
{\frac{W_{\Gamma_{LT}}^{(2)}}{2C_F}=-\int_{-T}^T dt_1\int_{-T}^T dt_2\; 
[D_{33}(0,2L,t_1+t_2,0)+D_{33}(0,0,t_1-t_2,0)]\\
-\int_{-L}^L dl_1\int_{-L}^L dl_2\; [D_{22}(0,l_1+l_2,2T,0)+
D_{22}(0,l_1-l_2,0,0)]\\
+4\int_{-T}^T dt\int_{-L}^L dl\; D_{23}(0,L+l,T-t,0).
}
This formula can be further simplified in the $T\to\infty$ limit since
various contributions are subleading.
Moreover, due to the identity (which holds since
$\Gamma_{LT}$ is a closed path)
\formula{transv.WL}
{\int_{\Gamma_{LT}}dx^\mu dy^\nu D_{\mu\nu}(x-y)=
\int_{\Gamma_{LT}}dx^\mu dy^\nu D^T_{\mu\nu}(x-y),
}
actually only the {\it transverse} part of the propagator contributes.
At zero mass $D_{\mu\nu}^T(p)$ is the same in all gauges and this is 
the reason why the final result is gauge-independent; however at
$\Lambda\neq0$ there is a dependence on the gauge vector $n_\mu$ coming
from the $\tilde D_{\Lambda,\mu\nu}$ part of 
the propagator.

A naive expectation would suggest that  $\tilde D_{\Lambda,\mu\nu}$
being proportional to $\Lambda^2$ gives a vanishing contribution to the 
Wilson loop. This is in general should not be taken for granted. 
Nevertheless, suppose
for a moment that this naive espectation is correct (this point
will be analyzed in detail in  next section).
Then one can consider only the  $\bar D_{\mu\nu}$ part of 
the propagator or even only
the transverse part $\bar D^T_{\mu\nu}$. This object is independent of the
gauge-fixing vector $n_\mu$ and therefore the Wilson loop is the same as
in covariant gauges. In particular, we can effectively
replace $\bar D_{\mu\nu}$ with $\delta_{\mu\nu}/(p^2+\Lambda^2)\equiv
\delta_{\mu\nu}D_\Lambda(p)$
because they have the same transverse part. If we divide by $2T$ in order
to eliminate the contributions subleading at $T\to\infty$
we obtain for the $O(g^2)$ analogous of the interquark potential
\formula{VL.def}
{V^{(2)}_\Lambda(2L)\equiv-
\lim_{T\to\infty}\frac{g^2 N_cW_{\Gamma_{LT}}^{(2)}}{4T}
}
the expression
\formula{VL.bar}
{V^{(2)}_\Lambda(2L)=
\frac{g^2N_cC_F}{2T}\int_{-T}^T dt_1\int_{-T}^T dt_2\; 
D_{\Lambda}(t_1+t_2,2L)+c}
where $c$ is the infinite renormalization constant 
\formula{c}
{c=\lim_{T\to\infty}
\frac{g^2 N_cC_F}{2T}\int_{-T}^T dt_1\int_{-T}^T dt_2\; D_{\Lambda}(t_1-t_2,0).
}
This is the well known ultraviolet divergence of point-like charges
and can be eliminated by fixing the potential to be zero
at $L\to\infty$. 
With simple manipulations involving the representation of the delta function
\formula{delta.rep}
{2\pi\delta(p_3)=\lim_{T\to\infty}\int_{-T}^Tdx_3\;
e^{ip_3 x_3}=\lim_{T\to\infty}
\frac{2\sin(p_3 T)}{p_3}
} and the
formula
\formula{large.T}
{\lim_{T\to\infty}\frac{(2\sin p_3 T)^2}{p_3^2}=
\lim_{T\to\infty}2T\cdot2\pi\delta(p_3),
}
from \rif{VL.bar} one obtains 
\formula{potential}
{V^{(2)}_\Lambda(2L)=g^2 N_cC_F\int_{\bar p}\frac{\exp(ip_2 \cdot 2L)}
{p^2+\Lambda^2}=g^2 N_cC_F\frac{\exp(-\Lambda\cdot 2L)}{4\pi\cdot 2L}.
}
This is the
formula of the screened Coulomb potential of two colored 
charges at distance $2L$
and reduces to the standard one when the infrared cutoff is removed.
We still stress that this formula is correct as far as we can 
neglect the contribution from the $\tilde D_\Lambda$ 
term in the limit $\Lambda\to0$.

\subsection{Vanishing of the $\tilde D_\Lambda$ term}

Now we prove that the $\tilde D_\Lambda$ term
of the propagator gives a vanishing contribution to the Wilson loop.
This is a non-trivial point since in this quantity
there is a problem of commutativity between
the limits $\Lambda\to0$ and $T\to\infty$.
This can be seen from the explicit expression of the
$\tilde D_\Lambda$ contribution to the interquark potential,
which up to an unessential infinite constant reads, working for definiteness
in the planar gauge,
\formulona{tilde.D.contrib}
{\tilde V^{(2)}_\Lambda(2L)=
\frac{g^2 N_cC_F}{2T}\int_{-T}^T dt_1\int_{-T}^T dt_2\; D_{33}(0,2L,t_1+t_2,0)\\
=-\frac{g^2 N_cC_F}{2T}\int_{p_E}
\frac{e^{ip_2\cdot 2L}\Lambda^2}{[p_3]^2(p_3^2+\bar p^2+\Lambda^2)}
\frac{(2\sin p_3 T)^2}{p_3^2}.
}
We see that we cannot blindly use the formula \rif{large.T} and compute
the integral directly at $T\to\infty$. We are instead forced to consider
finite $T$ and to make some subtle observation.
The idea is that the $p_3-$integral is dominated by the $p_3\simeq0$ 
region and that actually it reduces to an integral of the kind
\formula{an.cont}
{I_\alpha(T)=\int_{p_3} \frac{(2\sin p_3 T)^2}{(p_3)^{2+2\alpha}}
}
with $\alpha=1$. Strictly speaking this is a divergent integral; nevertheless
it can be defined in the region $-1/2<\alpha<1/2$ where its value is
\formulaX
{I_\alpha(T)=\frac2\pi\Gamma(-1-2\alpha)\sin(\alpha\pi)
(2T)^{1+2\alpha}\;.} 
After analytic continuation to $\alpha=1$ we obtain 
\formula{I1}
{I_1(T)=-\frac43 T^3.}
A simpler way to derive this formula is to derive equation \rif{I1} three 
times with respect to $T$: then the sinus representation of the
delta function is recovered and the integral is easily computed. In
other words we can use the formula, which has to be interpreted in the
sense of distributions,
\formula{T^3}
{\lim_{T\to\infty} \frac{(2\sin p_3 T)^2}{[p_3]^4}=
\lim_{T\to\infty}-\frac43T^3 \cdot 2\pi\delta(p_3).
}
Then the three-dimensional integration is as in \rif{potential}
and the final result is
\formulaX
{\tilde V^{(2)}_\Lambda(2L)=\frac{2g^2 N_cC_F}3 \Lambda^2T^2\;
\frac{\exp(-\Lambda\cdot 2L)}{4\pi\cdot 2L}\;.
}
We see that this contribution is quadratically
vanishing in the limit $\Lambda\to0$,
at {\it finite} $T$. The $T\to\infty$ limit {\it cannot} be taken before
the $\Lambda\to0$ limit.
This is the crucial point of our analysis.

For what concerns the situation with other gauges, we observe that in 
the light-cone gauge we have to compute exactly the same
integral as in \rif{tilde.D.contrib} and therefore we have the same
result. On the contrary, in the massive axial gauge we have to replace 
a factor $1/[p_3]^2$ with a factor
$(p_3^2+\Lambda^2)^{-1}$ and using \rif{fund} we see that the final result 
is {\it linearly} vanishing with $\Lambda T$. 
This means that the $O(g^2)$ Wilson loop 
test works for any gauge choice: but this is hardly a surprise since 
the $O(g^2)$ computation corresponds to the computation in an Abelian
theory. It is well known that problems begin in the non-Abelian
case and at higher orders
in perturbation theory, starting from order $O(g^4)$, therefore we cannot
conclude nothing about the final consistency of these Wilsonian gauge
choices at this level. Nevertheless we think that this analysis of the
free case is very instructive and allows to learn a lot about the possible
origin of problems. In particular we have learned that in an $O(g^4)$
computation the only part of the propagator we have to control is the
$\tilde D_\Lambda$ part, which is quite simple and surely can be studied with 
a relatively little analitical effort. In other Wilsonian 
formulations
of non-covariant gauges based on infrared cutoffs more
complicate than a mass-like term \cite{Litim,Geiger} the analysis 
is technically much more cumbersome but physically equivalent in what 
concerns the final results in the $\Lambda\to0$ limit. The drawback of
the generic cutoff is that the fine-tuning problem must be solved in order
to fix the correct boundary conditions (renormalization prescriptions)
such to have an infrared limit consistent with the gauge-symmetry;
this difficult problem is avoided with the mass cutoff since with this
choice the theory is 
Ward-identities-consistent to all scales. 
Finally we would stress the fact the planar gauge and in particular the
light-cone gauge
are expected to be much more regular than the axial gauge in the zero
mass limit: this expectation is reflected in the present computation by
the fact that the
$\tilde D_\Lambda$ term is quadratically suppressed in both planar and
light-cone gauge and merely linearly suppressed in axial gauge.

\section{Conclusions}

In \cite{paper.II} we pointed out that in algebraic non-covariant gauges
it is possible to build up a Wilsonian formulation of gauge theories
consistent with the Ward identities provided that the 
infrared cutoff is introduced as a formal ``mass'' term. However, we stressed
that this infrared cutoff cannot be physically interpreted
and {\it must} be removed in order to recover the
essential property of gauge-independence of physical quantities.
In this paper we have investigated the properties of the singular limit
$\Lambda\to0$ by explicitly computing various quantities. We have
seen in general that the pure axial gauge choice is
problematic since the Fourier transform of the propagator and
even the simplest Feynman diagrams which are finite in covariant
gauges are instead divergent at $\Lambda\to0$, for {\it any} configuration
of momenta.
Moreover, we argued that other gauge choices, namely the
planar gauge and the light-cone gauge are in a much better shape.
We have however seen that the structure of infrared divergences is quite subtle
even in these cases and that there are gauge-dependent quantities 
such as for {\it on-shell} configurations
of momenta the contribution from the gauge-dependent
term of the propagator $\tilde D_{\Lambda,\mu\nu}$
is not only non-vanishing, but even it is dominant with respect to the
standard contribution. Therefore the $\Lambda\to0$ limit for these quantities
is delicate. However these quantities are unphysical and 
this fact should not be considered as suggesting 
an inconsistency of the theory. The only way to prove the consistency
or the inconsistency of the approach is by considering true physical
quantities like the interquark potential as obtained from a Wilson loop
of size $2L\times 2T$ with $T\to\infty$. 
We have seen that the limits $\Lambda\to0$ and 
$T\to\infty$ must be studied with great care, since they do not commute.
In particular the crucial property of the exponentiation
of the Wilson loop is recovered only if the $\Lambda\to0$ limit is taken
{\em before} the $T\to\infty$ limit. We have seen that 
the planar gauge and the light-cone gauge appear to be 
much more regular than the axial gauge in the infrared limit. This is
explicitly realized in the Wilson loop $O(g^2)$ computation by the fact
that the gauge dependent term is quadratically suppressed at small
$\Lambda T$ whereas in axial gauge this term is only linearly suppressed.
We notice that in
covariant gauges, by accident, the order of limits is not crucial at
order $O(g^2)$, then this feature has not been recognized previously; 
nevertheless it is manifest in an $O(g^4)$ computation. 
The reason is that
the infrared cutoff breaks the BRST-invariance and therefore the cancellation
of the so called non-Abelian contributions to the Wilson loop 
no more works at $\Lambda\neq0$. Therefore the interquark
potential {\em cannot} be perturbatively defined at $\Lambda\neq0$ 
and $T\to\infty$, but only at {\em finite} $T$.
Finally we notice that the physical
interquark potential, defined exactly at $\Lambda=0$, is {\em independent} of 
the cutoff function choice when it is
computed order by order in perturbation theory.

For the future, we plan to study in detail what happens 
in the $O(g^4)$ computation. 
Actually, at finite $T$, we expect that in planar 
and light-cone gauges the computation reduces smoothly to the standard 
massless computation, which is consistent, whereas the axial gauge should
be very delicate and possibly inconsistent.
Still, we should notice that even in the cases where the $O(g^4)$
Wilson loop test fails, in principle this only indicate a failure of the
perturbative expansion and not necessarily of the full theory.
Nevertheless one would feel much more confortable with an approach
admitting a perturbative expansion. In such a perspective, the most
promising possibility seems to be the light-cone gauge which certainly 
deserves additional investigations.\\

{\it Note Added.} During the completion of this work we received a
communication from R. Soldati and A. Panza \cite{Soldati.Panza} 
who explicitly computed the
Wilson loop at order $O(g^4)$ in the massive axial gauge case and proved
that the $\Lambda\to0$ limit is {\em singular}. Therefore the axial gauge
choice seems to be definitely pathological even in the Wilson renormalization
group approach, at least at the perturbative level.

\vskip .5cm
{\large\bf Acknowledgements}\\

I acknowledge support from Padova University during the early stages
of this work. I thank R. Soldati and A. Panza for useful discussions
and for communicating to me the result of \cite{Soldati.Panza} before
publication.
\appendix

\section{Gauge-dependence of the pressure}

In this appendix we provide a very simple and illuminating example
which illustrates the gauge-dependence problem in presence of
a non-zero mass cutoff~$\Lambda$. 

Consider a free gas of photons in thermal equilibrium 
at temperature $T=1/\beta$ in a box of volume $V$. 
We can compute, as a typical 
quantity directly related to the partition function, 
the pressure of this gas in presence of the mass cutoff $\Lambda$. 
We obtain different result in different gauges, 
but all these results collapse to
the correct physical result in the physical limit $\Lambda\to~0$. 
This is trivial example, since the theory is free, nevertheless we
believe it is very instructive. We recall that the pressure is
a typical quantity which cannot be computed in perturbation theory in
thermal field theory and where the non-perturbative powerfulness of the
Wilson renormalization group approach could give an alternative
way of facing the problem. In this sense it is an interesting quantity.

We begin the computation by recalling some elementary facts (see for instance
\cite{Kapusta}).
In quantum field theory the thermodynamic pressure
\formula{def.P}
{p(\beta,V)=\frac1{\beta V}\ln Z_{\beta V}} 
is defined in terms of the partition function
\formula{Z.TV}
{Z_{\beta V}=\int[d\phi]
\exp\left(-\int_0^\beta
d\tau\int_V d^3x\;{\cal L}_E(\phi,\partial_\mu\phi;\Lambda_0)\right)
}
where ${\cal L}_E(\phi,\partial_\mu\phi;\Lambda_0)$ 
is the bare Euclidean lagrangian of the theory, which is a function of the
bare parameters depending on the ultraviolet cutoff $\Lambda_0$. For 
instance for a free scalar field of mass $\Lambda$ we have,
\formulaX
{{\cal L}_E(\phi,\partial_\mu\phi;\Lambda_0)=\left(\frac12\partial_\mu\phi
\partial^\mu\phi\right)_{\Lambda_0}+\frac12\Lambda^2\phi^2+
c_4(\Lambda,\Lambda_0),
}
where the notation 
$ \left(\frac12\partial_\mu\phi\partial^\mu\phi\right)_
{\Lambda_0}$ reminds that there is an
ultraviolet cutoff inserted in the propagator in momentum space and
the term $c_4(\Lambda,\Lambda_0)=\Lambda_0^4
[\tilde c_4+\tilde c_4'\Lambda^2/\Lambda_0^2+O(\Lambda^4/\Lambda_0^4)]$ 
is a vacuum energy counterterm 
of dimension four, 
which will be fixed later on by imposing the normalization condition  
\formula{norm.cond}
{\lim_{\beta\to\infty}\ln Z_{\beta V}=0,}
i.e. the partition function is fixed to be 1 at zero temperature.
It is interesting to notice that in order to impose this normalization
the ultraviolet regularization is needed even if the theory is free.
Using the gaussian integration formula we obtain
\formulaX
{Z_{\beta V}={\cal N}\det_{\beta V}(-\partial^2_E+\Lambda^2)^{-1/2}
\exp(-c_4(\Lambda,\Lambda_0)\beta V)}
where ${\cal N}$ is a temperature independent normalization factor to
be fixed later whereas the
determinant of the operator $-\partial^2_E+\Lambda^2$ (acting on functions
periodic in the imaginary time) is defined as
\formula{def.det}
{\det_{\beta V}(-\partial^2_E+\Lambda^2)^{-1/2}\equiv\exp
\left[-\frac12\Tr_{\beta V}\ln(-\partial^2_E+\Lambda^2)\right]
}
with
\formula{def.Tr}
{\Tr_{\beta V}\ln(-\partial^2_E+\Lambda^2)\equiv V
\sum_{n=-\infty}^\infty\int_{\vec p}
\ln[(p^4_n)^2+\vec p\;{}^2+\Lambda^2],
\quad p^4_n\equiv 2\pi n T.
}
Using the stardard summation formula\footnote{In order to use 
equation \rif{sum.formula} we have to remove 
the ultraviolet cutoff in the energy 
variable $p_4$, otherwise the summation reduces to a finite sum with $N\sim
\Lambda_0/T$ terms. In general the summation formula is defined up
to a possibly divergent constant which can be reabsorved in the vacuum
energy counterterm $c_4(\Lambda,\Lambda_0)$.
}
\begin{equation}\label{sum.formula}
\sum_{n=-\infty}^\infty\ln[(p^4_n)^2+X^2]=-2\log[1+n(X)]+
\beta X,
\end{equation}
where $n(X)=(\exp(\beta X)-1)^{-1}$ is the Bose-Einstein distribution
function, one obtains 
\formulaX
{\ln Z_{\beta V}=V\int_{\vec p}
\ln[1+n(\omega_p(\Lambda))]-
\frac12\beta V\int_{\vec p}^{\Lambda_0}\omega_p(\Lambda)-
\beta Vc_4+\ln{\cal N},}
with $\omega_p(\Lambda)=\sqrt{\vec p\;{}^2+\Lambda^2}$.
The first term vanishes at $\beta\to\infty$. Imposing
the normalization prescription \rif{norm.cond} fixes the counterterm
$c_4(\Lambda,\Lambda_0)$ and the normalization
factor ${\cal N}$ to be
\formulaX
{c_4(\Lambda,\Lambda_0)=-\frac12\int^{\Lambda_0}_{\vec p}\omega_p(\Lambda),
\quad{\cal N}=1,
}
therefore the pressure reduces to the well known expression
\formula{SB.law}
{p(T,\Lambda)=T\int_{\vec p}
\ln[1+n(\omega_p)]\equal{\Lambda\to0}\frac{\pi^2}{90}T^4.
}  
Now we can repeat the same computation in the gauge theory case which
is less trivial. We will
consider the general linear gauge \rif{lin.gauge} with $\xi_2\neq0$. 
Using the Gaussian integration formula we obtain
\formulaX
{Z_{\beta V}(L,\xi_2)=\N\det\xi_2^{-1/2}
\det_{\beta V}\partial^\mu L_\mu\det_{\beta V}D_{\mu\nu}^{1/2}}
where $\det\xi_2^{-1/2}$ comes from the integration of auxiliary fields,
$\det_{\beta V}\partial^\mu L_\mu$ from the integration of ghost fields
and $\det_{\beta V}D_{\mu\nu}^{1/2}$ from the integration of gauge
fields.
With a lengthy but straighforward computation, one obtains
the determinant of the Euclidean propagator in the general class of linear
gauges as
\formula{det.gen}
{\det D_{\Lambda,\mu\nu}(p)=\frac1{(p^2+\Lambda^2)^2}\frac{\xi_2}
{(p\cdot L)^2+\Lambda^2[L^2+\xi_2(p^2+\Lambda^2)]}.
}
Suppose for a moment that $\Lambda=0$: in this case
we see that the partition function is {\it gauge-independent} since
the $\xi_2$ term in the photon propagator determinant
is cancelled by the contribution coming from the 
integration of the auxiliary fields whereas
the $L-$dependent part is cancelled
by an analogous contribution from the ghost propagation, which in turn is
fixed by requiring BRST-invariance of the total action.
However, as can be explicitly seen,
this cancellation mechanism of the gauge-dependence does
not work in presence of an infrared regulator which breaks BRST-invariance.
Therefore at finite $\Lambda$ one obtains
an unphysical dependence on the gauge-fixing parameters.
In particular in the limit $\xi_2\to0$ i.e. strictly imposing 
the condition $L^\mu A_\mu=0$
one obtains for the logarithm of the partition function the explicit expression
\formulona{P.L}
{\ln Z_{\beta V}&=
2\Tr_{\beta V}\ln(-\partial^2_E+\Lambda^2)^{-1/2}+
\Tr_{\beta V}\ln((\partial\cdot L)^2+L^2\Lambda^2)^{-1/2}\\
&-\Tr_{\beta V}\ln((\partial\cdot L)^2)^{-1/2}-\beta Vc_4(\Lambda,\Lambda_0)+
\ln\N\;.
}
The first term is the expected one, corresponding to the pressure of 
two 
bosonic degrees of freedom; however there are also terms which are
explicitly $L-$dependent at $\Lambda\neq0$ and therefore
unphysical; the $L-$dependence only cancels at $\Lambda\to0$
where one recovers the correct result
\formula{P.phys}
{p(T;L)\equal{\Lambda\to0}\frac{\pi^2}{45} T^4\quad \forall\ L.
}
This is the generic situation, however
it is interesting to study what happens in specific gauge choices
since the gauge-dependence of the pressure can be less prononced than
expected. For instance in the light-cone gauge $L^2=n^2=0$ we
see that the $n_\mu-$dependent terms cancel even at $\Lambda\neq0$.
Actually the $n_\mu-$dependence cancels in the large class of non-covariant
gauges such as $L_\mu(p)$ does not depend on the $p_4$
variable (for instance this is the case for the planar gauge and the
Coulomb gauge): in this case the $L-$dependent terms in $\ln Z_{\beta V}$ 
have the form
$\beta V f(L,\Lambda,\Lambda_0)$ and therefore can be reabsorved
in the in the vacuum
energy counterterm $c_4(\Lambda,\Lambda_0)$. In other words, they are 
eliminated
by the normalization condition \rif{norm.cond} and do not contribute
to the pressure even at $\Lambda\neq0$. 
However the dependence on the quantization
direction should be expected in a two-loop computation; moreover it
is evident in other observables, for instance in the
Wilson loop computation, therefore in any case the
infrared cutoff $\Lambda$ {\it cannot} be physically interpreted as
a true mass term, consistently with the standard lore that the only
way to give a physical mass to a non-Abelian theory is via the
Higgs mechanism.
Nevertheless from this simple example one could extrapolate the conjecture 
(to be checked case by case perturbatively with
higher order computations or non-perturbatively with a numerical analysis) 
that for particular observables there are
classes of gauges more regular that others, 
in which the gauge-dependence is very mild even for non-zero $\Lambda$.
Clearly, this is a very interesting point in the spirit of phenomenological
numerical analysis and should be investigated
in the future.

\section{One-loop integrals}

In this appendix we give some generalities on the computation of 
Euclidean one-loop integrals in planar and light-cone gauges.
We begin by fixing our notations on integrals: for axial and
planar gauges we define
\formula{int.abbr.x}
{\int_x=\int d^4x,\quad\int_q=\int\frac{d^4q}{(2\pi)^4},\quad
\int_{q_3}=\int\frac{dq_3}{2\pi},\quad\int_{\bar q}=
\int\frac{d^3\bar q}{(2\pi)^3},
}
with $\bar q=(q^0,q^1,q^2)$, whereas for the light-cone gauge we define
\formula{int.abbr.q}
{\int_{q_\perp}=\int\frac{d^2q_\perp}{(2\pi)^2},\quad\int_{q_\parallel}=
\int\frac{d^2q_\parallel}{(2\pi)^2},
}
with $\vec q_\perp=(q^1,q^2)$, $\vec q_\parallel=(q^3,q^0)$.
Euclidean vectors are obtained after Wick rotation $p_4=ip_0$;
we shall use the notations
\formula{Euclidean}
{\!\!
p_E=(\vec p,i p_0),\quad q_E=(\vec q,i q_0),\quad p_E q_E
\equiv\delta_{\mu\nu}\ p_E^\mu\ q_E^\nu=-g_{\mu\nu}p^\mu q^\nu=-p q
}
and
\formulaX{p_\parallel^2=p_3^2,\quad p_\perp^2=\bar p^2\quad
\mbox{(planar gauge)}}  
\formulaX{p_\parallel^2=p_3^2+p_4^2,\quad p_\perp^2=p_1^2+p_2^2\quad
\mbox{(light-cone gauge)}
} 
In this appendix we will always work in Euclidean space even if for sake of
notational convenience the index $E$ will be neglected. 

For semplicity we will restrict 
our remarks to the computation of the one-loop self-energy of a scalar quark.
In general this is a rather complicate function $F=F(p_\parallel,
p_\perp,\Lambda)$, but we can give analytical estimations in the two limits 
$|p_\parallel^2|\ll p_\perp^2$ and $|p_\parallel^2|\gg p_\perp^2$. 
In the first case by putting $p_\parallel=0$ we see that the spurious
terms $1/[q\cdot n]$ cancel and then the one-loop integral can be put in 
an explicitly covariant form: therefore it can be computed
through the usual Feynman parametrization. The case 
$|p_\parallel^2|\gg p_\perp^2$ 
instead is more cumbersome, nevertheless it can be easily
implemented with a symbolic manipolation
package. As a matter of fact,
we prepared a set of routines based on the double Feynman parametrization
method allowing to compute analytically
all the integrals which are encountered in one-loop self-energy
diagrams, including finite parts, 
and we checked that the $\Lambda\to0$ limit
reproduces the known results of the standard approach.
However, for sake of brevity, here we simply sketch the analysis for
the simplest examples: the generalization to more complicate
cases is straighforward.

The general form of the Feynman integral one encounters in the evaluation of 
the quark self-energy (or its derivatives with respect to the mass) 
is given by the expression
\formula{general}
{F((p\cdot n)^2,p^2,\Lambda)=\int_q\frac{N(q,p,n)}
{[q\cdot n](q^2+\Lambda^2)((q+p)^2+m^2+\Lambda^2)^{1+\alpha}}.
}
The integral \rif{general} 
will be explicitly evaluated both in planar gauge and light-cone
gauge in next subsections.
\subsection{Planar gauge integrals}

Consider first the planar gauge case where $q\cdot n=q_3$.
The most convenient technical tool we found to manage this kind of integrals
is the double Feynman parametrization which consists in using the identities
\formula{F1}
{\frac1{q_3}\frac1{q^2+\Lambda^2}=\int_0^1dy\frac{q_3}
{[q_3^2+y(\bar q^2+\Lambda^2)]^2}}
and 
\formulona{F2}
{&\int_q\frac{N(q,p,n)}{(q^2+A^2)^a((q+p)^2+m^2+\Lambda^2)^b}=\\
&\frac1{B(a,b)}\int_q\int_0^1 dx\;
\frac{N(q-p(1-x),p,n)x^{a-1}(1-x)^{b-1}}
{[q_3^2+A x+(\bar q^2+p^2 x +m^2+\Lambda^2)(1-x)]^{a+b}.}
}
Then it is convenient to introduce the variable $z=(y-1)x+1$ 
lying in the interval $1-x\leq z\leq 1$ and to
rescale $\tilde q=z^{1/2} \bar q$; in this way equation \rif{general} can be
rewritten in the form
\formulaX
{F=\int_{q_3,\;\tilde q}\int_0^1dx\int_{1-x}^1dz\;
\frac{[q_3 N(q-p(1-x),p,n)]_{tr}(1-x)^\alpha}{z^{3/2}
[q_3^2+\tilde q^2+(p^2x+m^2)(1-x)+\Lambda^2 z]^{3+\alpha}},
}
where we have introduced the translated and rescaled numerator 
\formula{tr.axial}
{[q_3 N(q,p,n)]_{tr}=[q_3 N(q,p,n)]_{ q_3\to 
q_3-p_3(1-x)}^{\bar q^2=\tilde q^2/z}\;.
}
After symmetrization in $q_3$  and three-dimensional
angular average in the numerator, the momentum 
integrals can be performed by using the general formula 
\formula{int.1+3}
{\int_{q_3,\tilde q}\frac{(q_3^2)^{M_1}(\tilde q^2)^{M_2}}{(q_3^2+\tilde q^2
+A)^N}=\frac{B_1(M_1,M_2,N)}{8\pi^3A^{N-M_1-M_2-2}}
}
where $B_1(M_1,M_2,N)$ denotes the product of beta functions
\formula{BB}
{B_1=B(N-M_1-M_2-2,M_2+3/2)
B(N-M_1-1/2,M_1+1/2).}
The integral in
$z$ is trivial at $\Lambda=0$ and a little complicate at $\Lambda\neq0$
but still expressible in terms of elementary functions; the integral in 
$x$ instead in non-trivial. Nevertheless one can explicitly 
check that it is finite  and in general can be expressed
in terms of special functions. In the $\Lambda\to0$ limit the
analysis strongly simplifies and one obtain the esplicit result 
reported in the text.

With a simple generalization of this method one can compute Feynman integrals 
where the double pole $1/[q\cdot n]^2$ appears, without encountering 
any problem.
Moreover, the extension of this method to the integrals appearing
in the gluon self-energy computation is straightforward.

\subsection{Light-cone gauge integrals}

Consider now the light-cone gauge case where $q\cdot n=iq_4+q_3$.
Still the double Feynman parametrization can be used, it is enough
to replace identity \rif{F1} with
\formula{F1.LC}
{\frac1{iq_4+q_3}\frac1{q^2+\Lambda^2}=\int_0^1dy\frac{-iq_4+q_3}
{[q_4^2+q_3^2+y (q_\perp^2+\Lambda^2)]^2}.}
Using \rif{F2}, introducing the variable $z=(y-1)x+1$ and 
rescaling $\tilde q_\perp=z^{1/2} q_\perp$ the 
Feynman integral \rif{general} can be rewritten in the form
\formulaX
{F=\int_{q_\parallel,\;\tilde q_\perp}\int_0^1dx\int_{1-x}^1dz\;
\frac{[(-iq_4+q_3)N(q,p,n)]_{tr}(1-x)^\alpha}{z[q_\parallel^2+\tilde q_\perp^2+
(p^2x+m^2)(1-x)+\Lambda^2 z]^{3+\alpha}}
}
where we have defined the translated and rescaled numerator as
\formula{tr}
{[(-iq_4+q_3)N(q,p,n)]_{tr}=[(-iq_4+q_3)N(q,p,n)]_{\vec q_\parallel\to 
\vec q_\parallel-\vec p_\parallel(1-x)}^{q_\perp^2=\tilde q_\perp^2/z}.
}
Now we can perform the angular average in two dimensions
\formula{2d.ang.aver}
{<f(\vec q_\parallel\cdot \vec p_\parallel)>_{\Omega_2}\equiv
\frac1{2\pi}\int_0^{2\pi}d\theta\;f(q_\parallel p_\parallel\cos\theta),
}
by using the general formulae
\formula{2d.ang}
{<(\vec q_\parallel\cdot \vec p_\parallel)^{2n}>_{\Omega_2}=
\frac{\Gamma(n+1/2)}{\sqrt\pi\;\Gamma(n+1)}\;
q_\parallel^{2n}\;p_\parallel^{2n},\quad
<(\vec q_\parallel\cdot \vec p_\parallel)^{2n+1}>_{\Omega_2}=0.
}
In particular
\formula{2D.ang.sempl}
{<(\vec q_\parallel\cdot \vec p_\parallel)^2>_{\Omega_2}=
\frac12\;q_\parallel^2\;p_\parallel^2,\quad 
<(\vec q_\parallel\cdot \vec p_\parallel)^4>_{\Omega_2}=
\frac38\;q_\parallel^4\;p_\parallel^4.
} 
The integrals in $q_\parallel, \tilde q_\perp$ can be performed
by using the general formula
\formula{int.2+2}
{\int_{q_\parallel,\tilde q_\perp}\frac{(q_\parallel^2)^{M_1}
(\tilde q_\perp^2)^{M_2}}{(q_\parallel^2+\tilde q_\perp^2
+A)^N}=\frac{B_2(M_1,M_2,N)}{16\pi^2A^{N-M_1-M_2-2}}
} 
where
\formulaX
{B_2=B(N-M_1-M_2-2,M_2+1)B(N-M_1-1,M_1+1).
}
The integral in
$z$ is is trivial at $\Lambda=0$ and a little complicate at $\Lambda\neq0$
but still expressible in terms of logarithms of $\log(1-x)$; the 
integral in $x$ in non-trivial. Nevertheless one can explicitly 
check that it is finite and in general can be expressed
in terms of polylogarithmic functions. In the $\Lambda\to0$ limit and in 
the on-shell regime $p^2\to-m^2$ the
full expression strongly simplifies and we obtain the expression involving
logarithms given in the text.

With a simple generalization of this method one can compute Feynman integrals 
where the double pole appears, without encountering any problem.

Moreover, the extension of this method to the integrals appearing
in the gluon self-energy computation is straightforward.

\subsection{Feynman parameters integrals}

Finally we report some useful trick to perform integrals on Feynman 
parameters in particular limits. Suppose one has to compute 
the integral of some function
$f(x,\Lambda)$ which in the $\Lambda\to0$ limit is dominate from 
the $x\simeq0$ 
region; this is the case for instance for
\formula{typical}
{f(x,\Lambda)=\frac{P^\nu(x)}{(\Lambda^2+p^2 x)^n}}
where $P^\nu(x)$ is a polynomial a degree $\nu$ in $x$, with $\nu<n-1$. 
This is the typical 
expression we encounter in the computation of self-energy 
diagrams; then one can use the identity
\formulaX
{\int_0^1 dx\; f(x,\Lambda)=\int_0^\infty dx\; f(x,\Lambda)-
\int_1^\infty dx\; f(x,\Lambda)
}
and neglect the second contribution which is subleading at $\Lambda\to0$.
Then the first integral can be performed by using the formula
\formulaX
{\int_0^\infty dx\;
\frac{x^m}{(\Lambda^2+p^2 x)^n}=\frac{B(m+1,n-m-1)}{\Lambda^{2n}}
\left(\frac{\Lambda^2}{p^2}\right)^{m+1},
}
which holds for $m>-1$ and $n>m+1$. In the case in which the integral
is dominate by the $x\simeq1$ region it is sufficient to change the variable
$x'=1-x$ and use the same trick.

\end{document}